\documentclass[twocolumn,numberedappendix,twocolappendix,appendixfloats]{openjournal_dhayaa}

\usepackage{amsmath}
\usepackage{fontawesome5}
\usepackage{color}
\usepackage{xcolor}

\usepackage{natbib}
\setcitestyle{aysep={}}

\usepackage[colorlinks=true
  ,urlcolor=blue
  ,anchorcolor=blue
  ,citecolor=blue
  ,filecolor=blue
  ,linkcolor=blue
  ,menucolor=blue
  ,linktocpage=true
  ,pdfproducer=medialab
  ,pdfa=true
]{hyperref}

\newcommand{\Rvir}{R_{\rm vir}}
\newcommand{\Mvir}{M_{\rm vir}}
\newcommand{\Vvir}{V_{\rm vir}}
\newcommand{\cvir}{c_{\rm vir}}
\newcommand{\Mstar}{M_{\rm star}}
\newcommand{\Mgas}{M_{\rm gas}}

\newcommand{\eg}{{\sl e.g.}, }     
\newcommand{\dex}{{\,\,\rm dex}}

\newcommand{\msol}{\ensuremath{\, {\rm M}_\odot}}    
\newcommand{\msun}{\ensuremath{\, {\rm M}_\odot}} 
\newcommand{\kpc}{\ensuremath{\, {\rm kpc}}}         
\newcommand{\mpc}{\ensuremath{\, {\rm Mpc}}}

\newcommand{\sigmaDM}{\ensuremath{\sigma_{\rm3D}}}

\newcommand{\aform}{\ensuremath{a_{\rm form}}}
\newcommand{\spin}{\ensuremath{\lambda_{\rm h}}}

\definecolor{orcidlogocol}{HTML}{A6CE39}
\definecolor{purple}{RGB}{128, 0, 128}

\newcommand{\OrcidID}[1]{ \href[urlcolor = red]{https://orcid.org/#1}{\textcolor{lightgray}{\faOrcid}}}
\newcommand{\OrcidIDName}[2]{\href{https://orcid.org/#1}{#2}}

\defcitealias{Anbajagane2022BaryImprint}{A22}
\defcitealias{Shao2023Baryons}{S23}
\defcitealias{Singh2020CosmologyWithMORs}{S20}

\newcommand{\Shao}{Shao2023Baryons}
\newcommand{\Anba}{Anbajagane2022BaryImprint}
\newcommand{\Singh}{Singh2020CosmologyWithMORs}

\usepackage[T1]{fontenc}
\usepackage{ae,aecompl}

\begin{document}

\title{Baryonic Imprints on DM Halos: the concentration--mass relation and its dependence on halo and galaxy properties}

\author{\OrcidIDName{0000-0002-4561-7026}{Mufan (Jon) Shao}$^{1, \star}$}
\author{\OrcidIDName{0000-0003-3312-909X}{Dhayaa Anbajagane}$^{2, 3, \star}$}
\email{$^{\star}$mufanshao@gmail.com, dhayaa@uchicago.edu}
\affiliation{$^{1}$ Department of Physics, University of Chicago, Chicago, IL 60637, USA}
\affiliation{$^{2}$ Department of Astronomy and Astrophysics, University of Chicago, Chicago, IL 60637, USA}
\affiliation{$^{3}$ Kavli Institute for Cosmological Physics, University of Chicago, Chicago, IL 60637, USA\\}

\begin{abstract}
The halo concentration--mass relation has ubiquitous use in modeling the matter field for cosmological and astrophysical analyses, and including the imprints from galaxy formation physics is tantamount to its robust usage. Observational analyses, however, probe the matter around halos selected by a given halo/galaxy property --- rather than by halo mass --- and the imprints under each selection choice can be different. We employ the \textsc{Camels} simulation suite to quantify the astrophysics and cosmology dependence of the concentration--mass relation, $\cvir-\Mvir$, when selected on five properties: (i) velocity dispersion, (ii) formation time, (iii) halo spin, (iv) stellar mass, and (v) gas mass. We construct simulation-informed nonlinear models for all properties as a function of halo mass, redshift, and 6 cosmological/astrophysical parameters, with a mass range $\Mvir \in [10^{11}, 10^{14.5}] \msun/h$. There are many mass-dependent imprints in all halo properties, with clear connections across different properties and non-linear couplings between the parameters. Finally, we extract the $\cvir-\Mvir$ relation for subsamples of halos that have scattered above/below the mean property--$\Mvir$ relation for a chosen property. Selections on gas mass or stellar mass have a significant impact on the astrophysics/cosmology dependence of $\cvir$, while those on any of the other three properties have a significant (mild) impact on the cosmology (astrophysics) dependence. We show that ignoring such selection effects can lead to errors of $\approx 25\%$ in the baryon corrections of $\cvir$. Our nonlinear model for all properties is made publicly available.
\end{abstract}



\section{Introduction}

The halo concentration, which parameterizes the density profile of halos \citep{Navarro1997NFWProfile, Navarro2010EinastoVsNFW}, is a ubiquitous tool for modeling many astrophysical and cosmological observables: under the halo model prescription \citep{Cooray2002HaloModel}, the concentration--mass relation can be used, in conjunction with a model of the halo abundance and halo bias, to model the matter density field on quasi-linear and non-linear scales, and this has been verified through extensive validation of this model against predictions from N-body simulations which solve the full non-linear dynamics of the density field \citep[\eg][]{Mead2016HmCode}. Gaining the ability to model, and thus analyze, such scales allows us to extract significant information on cosmological/astrophysical processes \citep[\eg][]{Pandey2021DESxACT, Gatti2021DESxACT, Zacharegkas2021GGLensingDES, Miyatake2023HSCY3HaloModel}.

While N-body simulations precisely solve the gravitational dynamics of the matter density field \citep[see][for a recent review]{Angulo2022Sims}, they do not model the presence of baryons and the associated galaxy formation processes. Once such processes are included, the density field has qualitatively different behaviors on these quasi-linear and non-linear scales \citep[\eg][]{Chisari2018BaryonsPk, Schneider2019BaryonsPk, Giri2021Baryon} which then impacts small-scale measurements of observables connected to the density field, such as weak lensing \citep[\eg][]{Gatti2020Moments, Secco2022Y3Shear, Amon2022Baryons, Anbajagane2023CDFs}. The evolution of gas and stars in the halo significantly alters the halo density profile \citep[\eg][]{Duffy2010BaryonDmProfileDensity, Peirani2017DensityProfile}, and this evolution, in turn, depends on the rich physics of galaxy formation \citep[see][for a review]{Vogelsberger2020Hydro}. The downstream effects of such processes on the concentration--mass relation have been studied previously, and found to have strong mass- and redshift-dependence \citep{Duffy2010BaryonDmProfileDensity, Beltz-Mohrmann2021BaryonImpactTNG, Anbajagane2022BaryImprint, Shao2023Baryons}. The work of \citet[][henceforth \citetalias{\Shao}]{Shao2023Baryons} calibrated a part of these baryonic effects --- parameterized with four amplitudes corresponding to various processes associated with supernovae (SNe) and supermassive black holes (SMBH) --- using the \textsc{Camels} simulation suite \citep{Villaescusa-Navarro2021CAMELS}.

While these calibrated relations are already useful tools for theoretical modelling --- as alternative, baryonified relations to the fiducial N-body based concentration--mass relations \citep[\eg][]{Duffy2008Concentration, Bhattacharya2013Concentrations, Diemer2015Concentration, Diemer2019concentrations, Ishiyama2020UchuuConcentration} --- a potential missing piece in the robust application of these relations to data is the impact of halo selection effects on the model. While these selection effects are completely negligible for modelling the density field auto-correlations\footnote{The density field is the sum of matter contributions from all halos across the full range of masses, and thus there is no sense of a ``halo selection'' when working with only the density field.}, they become important when correlating the density field with any halo or galaxy field. For example, the correlation between galaxy positions and weak gravitational lensing (commonly denoted as galaxy-galaxy lensing or ``gg-lensing'') is a powerful tool in extracting cosmology from galaxy surveys \citep[\eg][]{Prat2012GGLY3, Lee2022GGLCmassDES, Miyatake2023HSCY3HaloModel, Dvornik2023GGLKiDs}. The galaxies observed in these surveys are not a fair sample of all possible galaxies in the Universe, but some specific subsample determined by survey configuration (\eg the wavelengths observed by the survey, the observing depth etc.), and requirements on the redshift estimation, survey completeness, and other systematic effects \citep{Porredon2021Maglim, Rodriguez-Monroy2022RedMagic, Bautista2021eBOSSCosmo}. An even simpler selection is a magnitude threshold (or a signal-to-noise threshold) --- which is always inherent in astronomical observations --- as it is fundamentally different from a halo mass threshold. Thus, it is apt to check whether these selections on halo properties other than mass result in a halo sample for which the baryon imprints look different from those of the halo mass-selected sample; an example of a concrete question is whether the concentration of stellar mass-selected halos depends differently on galaxy formation processes when compared to the concentration in a gas mass-selected sample.

This question can be directly probed by subsampling halo catalogs from cosmological hydrodynamical simulations to match the selections of a given survey \citep[\eg][]{Grandis2021Magneticum, Tamas2022SyntheticClusters, Seppi2023erositaTNG}, and then calibrate the baryon imprints on the concentration--mass relation for this specific subsample. In practice, the robust implementation of such selections is highly non-trivial and requires data products (such as the observed magnitudes of galaxies as they would be measured in a given survey) that are not easily available/created. Additionally, such an analysis would only be useful for the specific survey being considered given it is tuned for that specific survey's selection properties. In this work, we instead take a simpler and more general approach of splitting the halo sample according to various halo and galaxy\footnote{In this work, we denote halo properties as those that summarize the full matter distribution, i.e. including the dark matter, whereas galaxy properties include only the baryonic components.} properties and quantifying the differences in baryon imprints of the concentration--mass relation caused by this split. Such a choice minimizes the direct applicability of the calibrated relations to survey data, but in favor of enabling explorations of a broader range of selection effects and thus types of surveys (optical selection, X-ray selection etc.). The properties we choose to split the halo sample by are the: (i) total matter velocity dispersion, $\sigmaDM$, which is connected to X-ray luminosity \citep{Solinger1972XrayVel} (ii) halo formation time, $\aform$, which changes the star formation rate \citep[\eg][]{Conroy2009SFRTime, Hearin2013CAM}, and thus the galaxy colors (iii) the halo spin $\spin$, (iv) stellar mass, $\Mstar$, which is directly proportional to galaxy luminosity and (v) gas mass, $\Mgas$, which is directly proportional to luminosity in X-ray and millimeter wavelengths \citep[see][for a review on halo gas properties]{Kravtsov2012ClusterFormation}. These properties were chosen by looking at all available properties in the halo catalogs of the \textsc{Camels} simulations and selecting those connected to different observational selections.

\citetalias{\Shao} previously used the combination of 1000 \textsc{Camels} simulations to analyze the $\cvir-\Mvir$ relation of dark matter halos and the imprints of baryonic processes on it. That work built non-parameteric models of the halo concentration $\cvir(\Mvir, z, \Omega_{\rm m}, \sigma_8, A_{\rm SN1}, A_{\rm AGN1}, A_{\rm SN2}, A_{\rm AGN2})$ as a function of halo mass, redshift, two cosmological parameters, two supernovae feedback parameters, and two active galactic nuclei (AGN) feedback parameters. In this work we build upon those results by: (i) constructing models for the five additional properties described above --- $\sigmaDM$, $\spin$, $\aform$, $\Mstar$, and $\Mgas$ --- using the same simulations and methods as \citetalias{\Shao}, (ii) comparing mass-dependent features across all properties (the five above in conjunction with $\cvir$, as analyzed in \citetalias{\Shao}) to infer the origin of different baryon imprints, and; (iii) probing changes in the baryonic imprints on $\cvir$ due to subselecting halos by a given property.

Our work is organized as follows: in \S \ref{sec:DataMethods} we describe the \textsc{Camels} simulation data, including the properties we study, and the kernel-localized linear regression (\textsc{Kllr}) method used to analyze them. In \S \ref{sec:AstroCosmoDependence} we extract the full astrophysics and cosmology dependence of the property--mass relations for the five properties, and in \S \ref{sec:selectioneffect} we probe the baryon imprints on halo concentration for halos under different halo and galaxy selection effects, and quantify the error from ignoring such selections. We conclude in \S \ref{sec:conclusion}.

\section{Data and Methods}\label{sec:DataMethods}

\subsection{The \textsc{Camels} simulations}\label{sec:Camels}
The Cosmology and Astrophysics with MachinE Learning Simulations \citep[\textsc{Camels,}][]{Villaescusa-Navarro2021CAMELS}, is a simulation suite of $V = (25 \mpc/h)^3$ boxes, each with varying cosmological and astrophysical parameters. The cosmological parameters include the fractional matter energy density in the present epoch ($\Omega_{\rm m}$), and the root-mean-square of deviations in the linear density field smoothed on a scale of $8 \mpc/h$ ($\sigma_8$). The astrophysical parameters are $A_{\rm SN1}, A_{\rm SN2}$, which control the supernovae feedback, and $A_{\rm AGN1}, A_{\rm AGN2}$ which control the active galactic nuclei (AGN) feedback. The fiducial values of these parameters are $\Omega_{\rm m} = 0.3$, $\sigma_8 = 0.8$, and $A_{\rm SN1} =  A_{\rm SN2} = A_{\rm AGN1} =  A_{\rm AGN2} = 1$.

The \textsc{Camels} simulations were run with $256^3$ dark matter particles of mass $6.5\times 10^7 (\Omega_{\rm m} - \Omega_{\rm b})/0.251 \msol/h$, and then $256^3$ gas resolution elements with an initial mass of $1.27 \times 10^7 \msol/h$. The suite is specialized for training machine learning algorithms but the dataset can also be used for non-parameteric studies of astrophysics and cosmology. The full suite contains different types of runs, of which two are of interest to this work. The \textbf{Cosmic Variance (CV)} runs are 27 simulations run with the fiducial parameters quoted above and with varying initial conditions. The \textbf{1P} runs are 61 datasets (10 simulations per each of the 6 parameters and 1 fiducial run) that systematically vary a single parameter. All runs share the same initial conditions seed. The \textbf{Latin Hypercube (LH)} simulations are the main driver of this work. These are 1000 simulations, with varying initial conditions, that span a 6D parameter space of the input parameters of the simulation. The points in parameter space are chosen via an optimal scheme called a Latin Hypercube, which is a sampling technique that has become a frequent tool for building emulators for cosmology \citep{Heitmann2009Emulator}. The priors for the parameters are $0.1 < \Omega_{\rm m} < 0.5$, $0.6 <  \sigma_8 < 1.0$, $0.25 < A_{\rm SN1, AGN1} < 4$, and $0.5 < A_{\rm SN2, AGN2} < 2$.

The \textsc{Camels} simulation suite contains multiple galaxy formation models. In this work, we use the simulations from the galaxy formation model of \textsc{IllustrisTNG}, which we will interchangeably refer to as TNG, where the latter is a state-of-the-art high-resolution suite of simulations \citep{Nelson2018FirstBimodality, Pillepich2018FirstGalaxies, Springel2018FirstClustering, Marinacci2018FirstFields, Naiman2018FirstEuropium, Nelson2019TNG50, Pillepich2019TNG50}. The model choices made in the fiducial TNG simulations --- which are described in detail in \citet{ Weinberger2017Methods, Pillepich2018Methods} --- are adopted in the \textsc{Camels} simulations. The resolution level of the \textsc{Camels} simulation roughly corresponds to the TNG100-2 run.

There are four astrophysical parameters varied in \textsc{Camels}. The total energy injection rate (power) per unit star-formation, $e_w$, and the SN wind speed, $v_w$, are controlled by two normalization factors, $A_{\rm SN1}$ and $A_{\rm SN2}$, respectively. Thus $A_{\rm SN1}$ corresponds to the energy output rate, and $A_{\rm SN2}$ corresponds to the wind speed. The mass loading factor, $\eta$, which corresponds to the amount of mass transferred away through the SNe winds/feedback, is set by the combination $v_w^{-2}e_w$ and so is proportional to the combination $\eta \propto A_{\rm SN1}A_{\rm SN2}^{-2}$.

The supermassive black hole (SMBH) model has different modes of feedback --- a quasar/thermal mode, where the gas in the vicinity of the SMBH is thermally heated, and a radio/kinetic mode, where the gas is given a momentum kick instead. The kinetic mode is turned on during the phase of the SMBH where it has lower accretion rates, and is the dominant mechanism for halos at the Milky Way-scale and above \citep[\citetalias{Anbajagane2022BaryImprint}]{Weinberger2018SMBHsIllustrisTNG}. Note that the kinetic feedback is not continuously output into the halo. Rather, in the TNG model, the SMBH ``accumulates'' energy and expels it once a chosen energy scale is reached, which results in a ``bursty'' feedback process. The parameter $A_{\rm AGN1}$ controls the rate at which energy is accumulated by the SMBH and $A_{\rm AGN2}$ controls the minimum energy needed to initiate such a burst. Both AGN parameters control the rate of the bursts, while the latter also controls the speed of the kinetic feedback jets.

The uber halo sample used in this work --- which is obtained by combining halos from all 1000 LH simulations --- spans a mass range of $10^{11} \msun/h < \Mvir < 10^{14.5} \msun/h$; the minimum mass is set by requiring at least $\approx 1000$ particles in a halo, and the maximum mass is set by the number of available halos. We set the maximum $\Mvir$ of our analysis as the 10th most massive halo mass in the uber sample. The uber sample, as well as the mass range we use, follows the choices from \citetalias{\Shao}. Note that, since the maximum mass of a halo in a simulation depends strongly on $\Omega_{\rm m}$, and the value of $\Omega_{\rm m}$ is varied across the 1000 LH simulations, the high mass-regime of the model we build (see Section \ref{sec:KLLR}) is more heavily informed by simulations with a high $\Omega_{\rm m}$. The maximum halo mass in a single LH sim varies from a minimum of $10^{12}\msol/h$ to a maximum of $10^{14.7} \msol/h$ with a median of $10^{13.7} \msol/h$. This means our model predictions for $\Mvir > 10^{13.7}\msol/h$ are informed by half, or fewer, of the 1000 LH simulations.

In this work, we focus on six halo properties available in the halo catalogs as provided by  \textsc{Camels}, generated using the \textsc{Rockstar} pipeline \citep{Behroozi2013Rockstar}. The halo virial radius, $\Rvir$, is defined as the radius of the sphere containing an enclosed density that is $\rho = \Delta_{\rm vir}(z) \rho_c(z)$, where $\rho_c$ is the critical density at a given epoch, and $\Delta_{\rm vir}(z)$ is the density contrast described in \citet{Bryan1998vir}. The virial mass, $\Mvir$, is defined as the mass enclosed within this radius, $\Mvir = 4/3\pi\Rvir^3 \times \Delta_{\rm vir} \rho_c(z)$ and is the mass definition used throughout this work. The other halo properties we use in this work are
\begin{itemize}
    \item \textbf{Concentration} ($\cvir$), defined as $\cvir = \Rvir/r_s$, where $r_s$ is the scale radius defined in the NFW profile \citep{Navarro1997NFWProfile},
    \begin{equation}\label{eqn:nfw}
        \rho(r) = \frac{\rho_s}{\bigg(\frac{r}{r_s}\bigg)\bigg(1 + \frac{r}{r_s}\bigg)^2}
    \end{equation}
    and is computed in \textsc{Rockstar} through fits to the total matter density (not just DM density) profile of the halo. The profile is estimated in 50 equal-mass radial bins between $3\epsilon < r < \Rvir$, where $\epsilon = 2\kpc$ is the comoving force softening scale of the simulations \citep{Villaescusa-Navarro2021CAMELS}. The scale radius, $r_s$, is estimated from the maximum likelihood best-fit of Equation \ref{eqn:nfw} to the measured density profile. As discussed in \citetalias{\Shao}, prior works have shown the total matter distribution follows an NFW profile for $r > 0.1\Rvir$ but deviates from it below that length scale, primarily from the presence of the stellar component \citep[see their Figure 5]{Duffy2010BaryonDmProfileDensity}. Given the minimum fitting scale of $6\kpc$, the \textsc{Rockstar} concentration estimates will be influenced by this component. Thus, we treat these estimates as NFW-based approximations of the true density profile.

    \item \textbf{Velocity dispersion} ($\sigmaDM$), the mass-weighted standard deviation of all particle (DM, gas, and stars) velocities with the halo, with the different cartesian components added in quadrature to obtain the isotropic quantity. The isotropic velocity dispersion definition follows the definition of \citet{Yahil1977Velocity} and is larger than that of \textsc{Rockstar} by a simple, constant factor of $\sqrt{3}$. We implement this latter factor in our work. When presenting the $\sigmaDM - \Mvir$ relation, we normalize the dispersion by $\Vvir = \sqrt{G\Mvir/\Rvir}$ for visualization purposes.

    \item \textbf{Spin} ($\spin$), defined as the modulus of the halo angular momentum vector, estimated using all matter components, normalized by $\sqrt{2}\Mvir \Vvir \Rvir$ as defined in \citet[][see their Equation 5]{Bullock2001Spin}

    \item \textbf{Formation time} ($\aform$), the scale factor at which the halo first achieved half of its present day mass, $\Mvir(a = \aform) = 0.5 \Mvir(a = 1)$. The halo merger tree, from which this quantity is derived, is estimated using \textsc{ConsistentTrees} \citep{Behroozi2013ConsistentTrees}.

    \item \textbf{Stellar mass} ($\Mstar$), the total mass of stars within $\Rvir$.

    \item \textbf{Gas mass} ($\Mstar$), the total mass of gas within $\Rvir$. The quantity uses all gas and has no temperature cuts.
\end{itemize}

\subsection{Kernel-localized linear regression}\label{sec:KLLR}

Many halo property scaling relations are adequately fit with a simple power law, i.e. a log-linear relation \citep[\eg][]{Evrard2008VirialScaling, Allen2011CosmoClusterReview, Kravtsov2012ClusterFormation, Kravtsov2018SMHMRelation, Lim2021GasMass, Lee2022rSZ}, while some are fit with extended parametric models such as a broken power-law model with two power-law indices instead of one \citep[][see their Figure 2 and references therein]{Wechsler2018GalaxyHaloConnection}. These parameterizations in general need not capture all the non-linearity in a scaling relation. The use of simple power-laws to represent a scaling relation is motivated by the observed behaviors in gravity-only simulations, where there are only a few unique scales that can break a self-similar scaling; a self-similar scaling would manifest in the form of a constant, scale-independent power law index. The inclusion of galaxy formation introduces many unique mass/time-scales \citep[\eg][]{Pillepich2018Methods, Weinberger2017Methods, Vogelsberger2020Hydro} and thereby causes scaling relations for most halo properties to deviate from the power law behavior \citepalias{Anbajagane2022BaryImprint}. With the surplus of simulation data available to us, we can now explore data-driven models that accurately capture non-linearities with halo mass without assuming fixed functional forms such as scale-independent power-laws or broken power-laws.

One approach to building model-independent representations of scaling relations is machine learning (ML) techniques, such as random forest, neural networks, and more \citep[\eg][]{Machado2020GasShapesSHAP, Stiskalek2022ML}. Such methods are also more difficult to interpret, and thereby more difficult to extract physical explanations from. Though there are approaches that can alleviate the interpretability issue \citep[\eg][henceforth \citetalias{\Anba}]{Ntampaka2019XrayClustersML, Machado2020GasShapesSHAP, \Anba}. An alternative then is to continue with well-known, well-understood linear models but extend the models to capture some or most of the non-linearity.

Kernel-localized linear regression \citep*[\textsc{Kllr},][]{Farahi2022KLLR} is an extension that allows such flexibility. The input data (i.e. halos) are weighted according to a kernel, which is chosen to be a Gaussian kernel in $\log_{10}\Mvir$, and a linear regression is fit to the kernel-weighted data. As the kernel is shifted in the log-mass direction, the method measures the non-linear scaling relation using multiple, connected linear pieces. \textsc{Kllr} has had extensive use in extracting nonlinearity in different halo property--mass scaling relations \citep[][\citetalias{\Anba}, \citetalias{\Shao}]{Farahi2018BAHAMAS, Anbajagane2020StellarStatistics, Anbajagane2022GalVelBias}. In this work, we continue on the efforts of \citetalias{\Shao} and use \textsc{Kllr} to study a multivariate scaling relation that is still modelled as being locally linear in halo mass. 

The locally linear, multivariate scaling relation of a halo property is written in \textsc{Kllr} as,
\begin{align}\label{eqn:Kllr}
    \log_{10}Y = &\,\, \pi(\Mvir, z) + \alpha_M(\Mvir, z) \log_{10}\Mvir\nonumber\\
    & + \sum_X\alpha_X(\Mvir, z) \log_{10}X,
\end{align}
where $Y \in \{\cvir, \sigmaDM, \aform, \spin, \Mstar, \Mgas\}$ is the target property to be predicted, $\pi$ is the intercept, $\alpha_M$ is the slope with halo mass, $\alpha_X$ is the slope with other parameters, which are $X \in \{\Omega_{\rm m}, \sigma_8, A_{\rm SN1}, A_{\rm AGN1}, A_{\rm SN2}, A_{\rm AGN2}\}$. The slopes and intercepts are all functions of halo mass and redshift. The property, $Y$, is thus modelled as a function of mass, redshift, and 6 cosmological/astrophysical parameters. The dependence on the cosmological/astrophysical parameters is also a function of mass. When analyzing an individual, single simulation (i.e the individual 1P simulations), we drop all terms with $\alpha_X$ --- as a single simulation has a singular, fixed value for the cosmlogical and astrophysical parameters --- and only keep the slope with mass. All analyses with the LH simulations use all terms in Equation \eqref{eqn:Kllr}, including the slopes $\alpha_X$. In all analyses of the LH sims that follow, the set of parameters in $X$ is the same as that denoted above. In Section \ref{sec:selectioneffect} below, we will also discuss extensions to the model described above, where the slopes and intercepts are functions of other halo properties as well.

We compute a \textsc{Kllr} model for each of the 34 available snapshots in $\textsc{Camels}$, using the uber sample at each redshift between $0 < z < 6$. In practice, this means we compute Equation \eqref{eqn:Kllr} independently for the halo sample from each snapshot. Our results in this work will focus on the model for just $z = 0$ but models for the full range of redshifts are provided publicly. We compute the scaling relations for every $0.1 \dex$ in mass, with the minimum mass set at $10^{11} \msun/h$ and a maximum mass set by the 10th most massive halo of the uber sample. The upper limit choice excludes regions where the model solutions are imprecise given the low number of halos. The \textsc{Kllr} technique uses a Gaussian kernel in $\log_{10}\Mvir$, with a width of $0.3 \dex$. Uncertainties on all the \textsc{Kllr} parameters of equation \eqref{eqn:Kllr} are obtained by estimating regression coefficients for 100 bootstrap realizations of the data. Thus, we have 100 versions of equation \eqref{eqn:Kllr} that then quantify the uncertainty in the predictions.

Other techniques, such as Gaussian processes, can also be used to capture the non-linearity of a relation in a non-parametric way with minima assumptions. For example, \textsc{Kllr} and Gaussian processes have been previously used to extract the same non-linear relations in halo quantities \citep[\eg][]{Farahi2021PoPE, Farahi2022ProfileCorr}. In this work, we continue using \textsc{Kllr} given its prior, extensive use in characterizing halo scaling relations.

\section{Dependence on astrophysics \& cosmology}\label{sec:AstroCosmoDependence}

We detail the baryon imprints on each halo property in the following sections. The halo concentration is omitted, as the results have already been presented in \citetalias{\Shao}. We still show the slopes of $\cvir-\Mvir$ in Figure \ref{fig:TNGslope}, which illustrate the main findings from \citetalias{\Shao} that $\cvir$ is highly sensitive (insensitive) to the SN (AGN) feedback parameters, with a clear non-linear dependence on $\Mvir$. All results in this section come from our \textsc{Kllr} model fits, and uncertainties are 68\% confidence intervals obtained from 100 bootstrap resamplings.

\subsection{Particle velocity dispersion, $\sigmaDM$}\label{sec:sigmaDM}

\begin{figure*}
    \centering
    \includegraphics[width = 2\columnwidth]{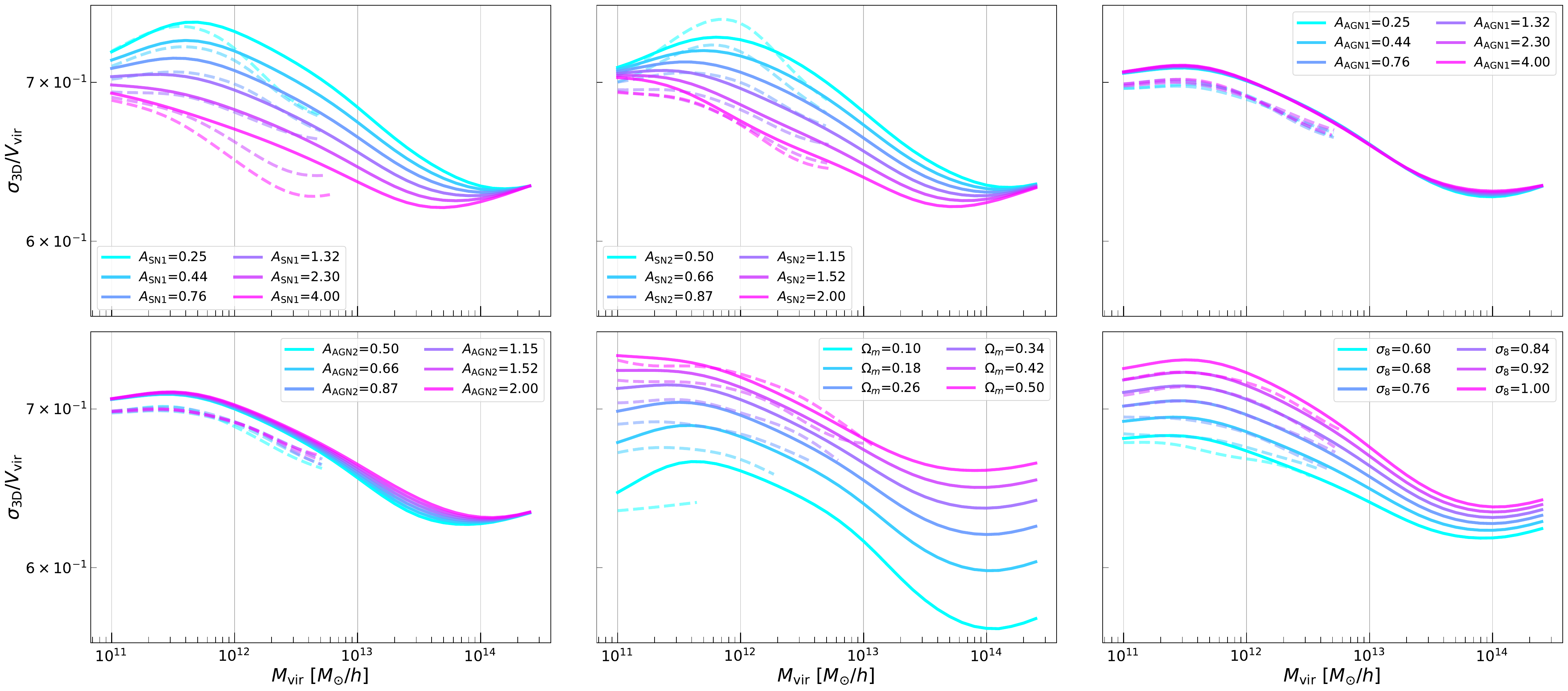}
    \caption{The \textsc{Kllr} $\sigmaDM-\Mvir$ scaling relation. We normalize $\sigmaDM$ by $\Vvir$, so the variation across mass can be more easily seen. The solid lines are the prediction from our model, made using the 1000 LH simulations. We fix the input parameters to their fiducial values, and then vary just a single parameter at a time in each panel. There is a rich variety of trends in how the halo property responds to changes in both the astrophysical and cosmological parameters. The dashed lines are the \textsc{Kllr} result from the 1P simulation, and our model captures the trends in the 1P runs. In all cases, the LH result and its associated 1P result correspond to the same set of simulation parameters ($\Omega_{\rm m}, \sigma_8, A_{\rm SN1/2,\,AGN1/2}$). The match between the 1P and LH results is degraded by impact of cosmic variance on the former; see Section \ref{sec:sigmaDM} for details. The offsets between our model and the 1P scaling relations are generally $1-2\%$, while the $1\sigma$ from cosmic variance is $1\%$. Each 1P result is cut at the $\Mvir$ corresponding to the 10th most massive halo in the simulation. This cutoff varies significantly with $\Omega_{\rm m}$, which alters the number of halos in the simulation, and is mostly constant with other halos. The high $\Mvir$ predictions of the LH-based model are more strongly informed by simulations with high $\Omega_{\rm m}$; see Section \ref{sec:Camels}.}
    \label{fig:TNGsigmaDM}
\end{figure*}

Figure \ref{fig:TNGsigmaDM} shows that the slope of $\sigmaDM$ with $\Mvir$ is a strongly non-monotic function of mass, consistent with \citetalias{\Anba} and with \citet{Anbajagane2022GalVelBias} who showed this behavior for a variety of simulation models. The velocity dispersion increases with $\Omega_{\rm m}$ and $\sigma_8$ across all masses, and the influence of $\sigma_8$ (as determined by the amplitude of the slope between $\sigmaDM - \sigma_8$, shown in Figure \ref{fig:TNGslope}) is greatest, compared to that of $\sigmaDM$ with the other five simulation parameters, for most of the mass range before dropping by a factor of 3 at $\Mvir \sim 10^{14} \msol$ (see Figure \ref{fig:TNGslope}).

In Figure \ref{fig:TNGsigmaDM}, we have also compared the model fit using the 1000 LH sims (solid line) to fits from individual 1P simulations (dashed lines). The dashed lines are a subset of six simulations from the available eleven 1P simulations. Each dashed line extends until the mass of the 10th most massive halo in the simulation. For each 1P simulation, we have first estimated the property--mass scaling relation using \textsc{Kllr} and then also predicted the relation with the LH-based model, by inputting the values of simulation parameters ($\Omega_{\rm m}, \sigma_8, A_{\rm SN1}, A_{\rm AGN1}, A_{\rm SN2}, A_{\rm AGN2}$) corresponding to that 1P simulation. Figure \ref{fig:TNGsigmaDM} shows that the model captures the mass-dependent trends of the $\sigmaDM-\Mvir$ relation with all six simulation parameters. However, there are also clear offsets between the 1P results and the model predictions.

There are two reasons such offsets will arise: (i) The first is that each dashed curve is estimated using a single 1P simulation and is thus significantly affected by cosmic variance given the small volume of each simulation. The model parameters, on the other hand, are estimated using 1000 LH simulations and are therefore much less impacted by cosmic variance. Note that since all 1P simulations share the same initial conditions, the resulting curves will all have correlated offsets, induced by cosmic variance, from the true mean relation. Thus, multiple dashed lines can exhibit coherent offsets even in the case where the model is perfect.\footnote{Note that even though the scaling relations estimated from the 1P simulations suffer from cosmic variance, the relative differences between the relations are cosmic variance-suppressed due to all 1P runs using the same initial conditions. Thus, these relative differences are still useful in verifying that our LH-based \textsc{Kllr} model is capturing the variations in the relations arising from varying each of the six simulation parameters' values.} (ii) The second is that the \textsc{Kllr} model we choose assumes a linear relation between the parameters, at fixed mass. \citetalias[][]{\Shao}, in their study of the $\cvir -\Mvir$ relation, verified that including higher order terms in Equation \eqref{eqn:Kllr}, namely quadratic terms such as $(\log_{10}X)^2$, results in only a minor improvement in the accuracy of the model, and we have verified the same for the relations studied in this work. We prioritize limiting our model to linear terms in $\log_{10}X$ as it simplifies the model and enhances its interpretability.

We have furthermore checked that, in general, a large fraction --- and in many cases all --- of each offset can be accounted for via cosmic variance, where the latter is estimated by measuring the property--mass mean relation on each of the 27 CV runs in the \textsc{Camels} suite and estimating the 68\% confidence interval of these 27 scaling relations. We quote the characteristic cosmic variance for each property in the figure captions, alongside the characteristic values of the offsets, which are below 5\% for the halo properties and rise to 10-20\% for the galaxy properties. In general, all offsets are within the $95\%$ confidence interval defined by cosmic variance. The offset amplitudes are also generally subdominant to the change in the model predictions as we vary the simulation parameters, and thus all qualitative discussions in this work are unaffected by the presence of these offsets.

Other works have studied the cosmology dependence of this relation. \citet[][henceforth \citetalias{\Singh}]{Singh2020CosmologyWithMORs} used 15 simulations from the \textsc{Magneticum} suite \citep{Hirschmann2014MGTM} that span a variety of cosmological parameters and found that $\sigmaDM$ is independent of $\sigma_8$; in particular, they determine the \textit{mass-independent} slope between $\sigmaDM$ and $\sigma_8$ to be $\alpha_{\rm \sigma_8} < 0.01$ at 95\% confidence (see their Table 5). Our results indicate the \textit{mass-dependent} slope drops towards $\alpha_{\rm \sigma_8} \approx 0.05$ at $\Mvir = 10^{14} \msol/h$ and suggests the slope decreases further towards higher mass. The mass sample in \citetalias{\Singh} extends to $\Mvir = 10^{15.5}\msol$ (about $1\dex$ higher than our sample). Therefore, we expect their estimate of the mass-independent slope to be lower due to the presence of these massive halos in their sample. \citetalias{\Singh} also show that the the slope between $\sigmaDM$ and the baryon fraction $f_b = \Omega_{\rm b}/\Omega_{\rm m}$ is $\alpha_{ f_{\rm b}} = -0.05\pm 0.01$. The \textsc{Camels} suite varies $\Omega_{\rm m}$ while fixing $\Omega_{\rm b} = 0.049$, which means changes in $\Omega_{\rm m}$ can be directly translated to those in $f_{\rm b}$. The slopes of $\sigmaDM$ with these two properties are inter-related through a simple negative sign; $\alpha_{\rm \Omega_{\rm m}} = -\alpha_{\rm f_{\rm b}}$. We find that at high mass, our slope estimates are higher than those of \citetalias{\Singh}, with our estimates nearing $-\alpha_{\rm \Omega_m} = -0.1$, but are consistent with their work when considering the behavior of halos with lower masses.

Turning to the SNe feedback parameters, $A_{\rm SN1}$ and $A_{\rm SN2}$, we see $\sigmaDM$ scales inversely with their amplitude. This is expected as SNe feedback causes the halo matter distribution to become more disperse \citep[][\citetalias{\Anba}, \citetalias{\Shao}]{Bryan2013BaryonImpactOnShapes}, which in turn will lower $\sigmaDM$ due to the strong correlation between the $\sigmaDM$ and $\cvir$ \citepalias{\Anba}. Weakening SNe feedback causes the maximum and minimum value of $\sigmaDM/\Vvir$ to shift to higher masses, and vice-verse when strengthening the feedback.
\citet{Lau2010BaryonDisspiation} compare the velocity dispersion (of dark matter) in hydrodynamic simulations that are either non-radiative or contain gas cooling plus star formation. They find that the presence of cooling and star formation increase $\sigmaDM$ (see their Figure 1) and this is due to the adiabatic contraction of the halo potential enabled by the radiative processes \citep{Blumenthal1986AdiabaticContraction, Gnedin2004AdiabaticContraction}. Increasing the amplitude of SNe feedback hinders this cooling process, and will reduce $\sigmaDM$, as observed in Figure \ref{fig:TNGsigmaDM}.

Finally, the AGN parameters have a weak impact on this relation, as expected from their weak impact on the halo concentration \citepalias{\Shao}. The $A_{\rm AGN1}$ parameter has no significant impact across the whole mass range, while $A_{\rm AGN2}$ influences the relation above Milky way (MW) masses. At these larger mass-scales, the thermal feedback of the AGN becomes less dominant, and the kinetic feedback takes over \citepalias[][see their Figure 7]{Anbajagane2022BaryImprint}. $A_{\rm AGN2}$ controls the energy of the kinetic bursts, and so its strongest impact is expected to be in this mass range. Unlike with the SNe parameters, this AGN parameter \textit{increases} $\sigmaDM$. However, this is expected as $A_{\rm AGN2}$ is directly proportional to the energy (and thus momentum) input into the gas. Stronger bursts result in more significant momentum increases, which will then increase the velocity dispersion within the halo. See \citet{Weinberger2017Methods} for details on the AGN implementation in \textsc{IllustrisTNG}. Note that these results \textit{do not} imply that the AGN as a whole has only a mild impact on the halo internal dynamics. Instead, it implies the \textit{specific AGN processes} being varied in \textsc{Camels} --- as parameterized by $A_{\rm AGN1}$ and $A_{\rm AGN2}$ --- have a mild impact.

\subsection{Formation time, $\aform$}

\begin{figure*}[!ht]
    \label{sec:aform}
    \centering
    \includegraphics[width = 2\columnwidth]{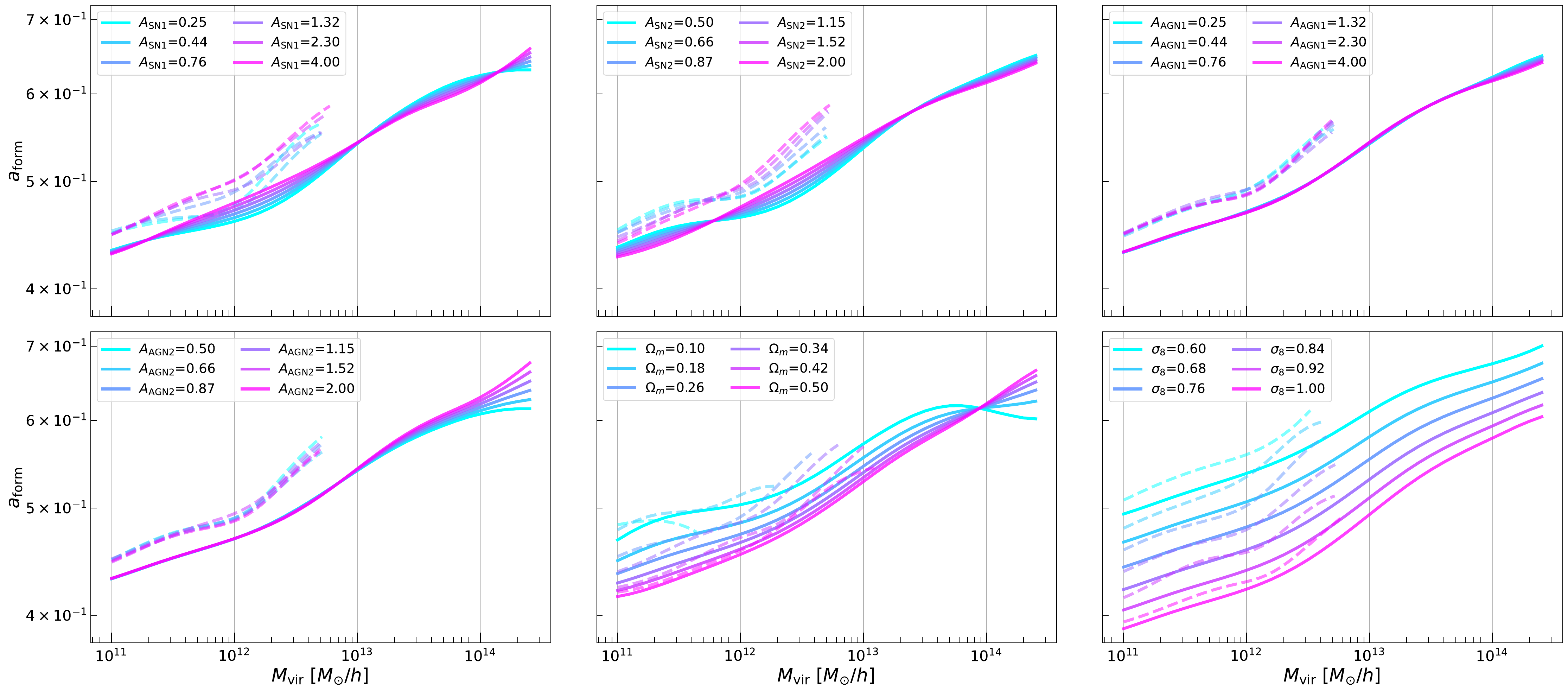}
    \caption{Same as Figure~\ref{fig:TNGsigmaDM} but for the $\aform-\Mvir$ scaling relation. The offsets between our model and the 1P scaling relations are generally $\approx5\%$ while the $1\sigma$ from cosmic variance is $3\%$.}
    \label{fig:TNGaform}
\end{figure*}

Figure \ref{fig:TNGaform} shows the scaling relation for $\aform$ with $\Mvir$, which is mostly log-linear/monotonic except for variations induced by changing the SNe feedback amplitudes. The shape of the fiducial relation (best seen in the top right panel) is log-linear with a deviation at MW masses also seen in \citetalias[][see their Figure 4]{\Anba}. Larger values of $\aform$ indicate halos formed \textit{later} in time, and the quantity is limited to the range $0 \leq \aform \leq 1$.

First, we discuss the cosmology dependence, where $\aform$ decreases with $\sigma_8$ and $\Omega_{\rm m}$. The slope between $\aform$ and the former is nearly a factor of $6$ larger than the slopes with all other simulation parameters (see Figure \ref{fig:TNGslope}). It is nearly constant below MW masses and increases with mass for halos above MW masses. The slope with $\Omega_{\rm m}$ increases over the whole mass-range, from $-0.1$ at small masses to $0$ at $10^{14}\msol/h$, and increasing to positive values at higher masses. At the highest masses, the positive slope is found to be non-zero at $2\sigma$. If this is a robust feature, it may arise from a nonlinear coupling between $\Omega_{\rm m}$ and other simulation parameters; such nonlinear coupling has been found in other analyses of \textsc{Camels}, though in the context of the astrophysical parameters \citep{Gebhardt2023CamelsAGNSN} and we discuss other such couplings below. In general, we expect $\aform$ to decrease with $\sigma_8$ and $\Omega_{\rm m}$ as any increases in these parameters' amplitudes (which change the amount of matter and amount of clustering of this matter) make it easier for structures to form, which then causes their formation to happen earlier and in turn cause $\aform$ to decrease \citep{Gao2005AgeDependence, Li2008FormationTimes}.

The SNe feedback parameters have a rich variety of impacts on $\aform$. At lower mass, $\Mvir < 10^{11.5} \msol/h$, they either have no influence or they reduce $\aform$. The direction of this change is expected as in small halos with shallow gravitational potentials, SNe feedback can blow out the gas \citep{Muratov2015GasOutflowsFIRE}. The halo will then require more time to grow and achieve its present-day mass. At higher masses, $10^{11.5} \msol/h < \Mvir < 10^{13} \msol/h$ this behavior qualitatively changes as increasing SNe feedback causes halos to form \textit{earlier}. This is an example of the non-linear coupling between the many astrophysical processes that occur within a halo, and showcases how varying a single parameter can change the impact of multiple processes in the halo rather than the impact of just the single process associated with that parameter.

In this case, the behavior is consistent with a non-linear coupling between the SNe feedback and the AGN feedback, as discussed by \citet{Gebhardt2023CamelsAGNSN} in their analysis of the same \textsc{Camels} simulations we use here. They show that for strong SNe feedback, the gas in these more massive halos is dispersed into the further outskirts of the halo. The SNe are generally not strong enough to eject the gas out of the halo, so the total mass within the halo remains conserved, while the distribution of matter within it changes (\eg see results on $\cvir$ from \citetalias{\Shao}). When AGN feedback is a prominent process in the halo, which begins at $\Mvir \approx 10^{11.5}\msol/h$ and is maximized at $\Mvir \approx 10^{12.5}\msol/h$ (see \citetalias{\Anba}, their Figure 7), it can eject gas far outside the halo boundary and thereby alter the halo mass distribution. However, the impact of the AGN jets depends crucially on the amount of gas present in the vicinity of the AGN. If the gas is more dispersed before the AGN turns on, then the impact of the AGN on the halo is weaker. Since SNe affect this dispersion of gas, they also control the impact of the AGN feedback, as shown in \citet{Gebhardt2023CamelsAGNSN}.\footnote{Similar arguments have also been used to interpret the negative correlation of the halo concentration (which denotes how much material is near the halo center, and thus the AGN) with gas properties \citep[\eg][]{Boryana2023SZ}.}

Thus, increased SNe feedback enhances the dispersion of the gas and reduces the impact of ejection from AGN jets. Halos will then retain more of their gas and require less time to achieve their present-day mass. They can therby form later in cosmic history. This causal picture is corroborated by the $\Mgas$ behavior detailed in Figure \ref{fig:TNGMgas} below. Note that the conventional behavior of increased SNe feedback reducing $\aform$ is recovered at galaxy group/cluster mass-scales, $\Mvir > 10^{13.5} \msol/h$. Note also that $A_{\rm SN1}$ has a further cross-over point at $\Mvir \approx 10^{14}\msol/h$, which is within a factor of 2 of the mass-scale of a similar crossover point found in the $\Omega_{\rm m}$ dependence. This suggests the $\Omega_{\rm m}$ and $A_{\rm SN1}$ may be linked via a non-linear coupling, similar to that of the SNe and AGN feedback. A larger halo sample with better statistical power is required to verify this.

The AGN feedback has very little influence on the $\aform-\Mvir$ relation. The only notable dependence is on $A_{\rm AGN2}$ at the high mass end, measured to be non-zero at $3\sigma$, and this implies that increasing $A_{\rm AGN2}$ leads to halos needing \textit{less} time to achieve their present-day mass. Note that $A_{\rm AGN2}$ does not affect the total energy injected into the halo, and instead affects the energy in each \textit{burst}. Thus, the parameter impacts the ``burstiness'' or stochasticity of the AGN jet feedback. In Figure~\ref{fig:TNGMgas} below we show that increasing $A_{\rm AGN2}$ \textit{reduces} the gas mass in the halo at the present time. This reduction will then change the corresponding estimate of $\aform$ to be higher. As a pedagogical example, consider a halo of $\Mvir = 10^{14.5}\msol/h$ at $z = 0$ with a given mass accretion history $\Mvir(a)$. The formation time is defined as the scale factor where $\Mvir(\aform) = 0.5\Mvir(a = 1)$. Thus, a process that is most impactful at $a \sim 1$ will affect the present-day mass and thereby affect the $\aform$ estimate. If the process in question is a large-scale ejection event, such as one generated by AGN feedback, then $\Mvir(a = 1)$ is reduced while the overall history $\Mvir(a)$ is left unaffected, especially for the earlier epochs. In this case, $\aform$ is reduced as the mass to be achieved ($\Mvir$ at $a = 1$) has been reduced, and so the halo can achieve it at earlier times.

\subsection{Spin, $\spin$}\label{sec:spin}

\begin{figure*}
    \centering
    \includegraphics[width = 2\columnwidth]{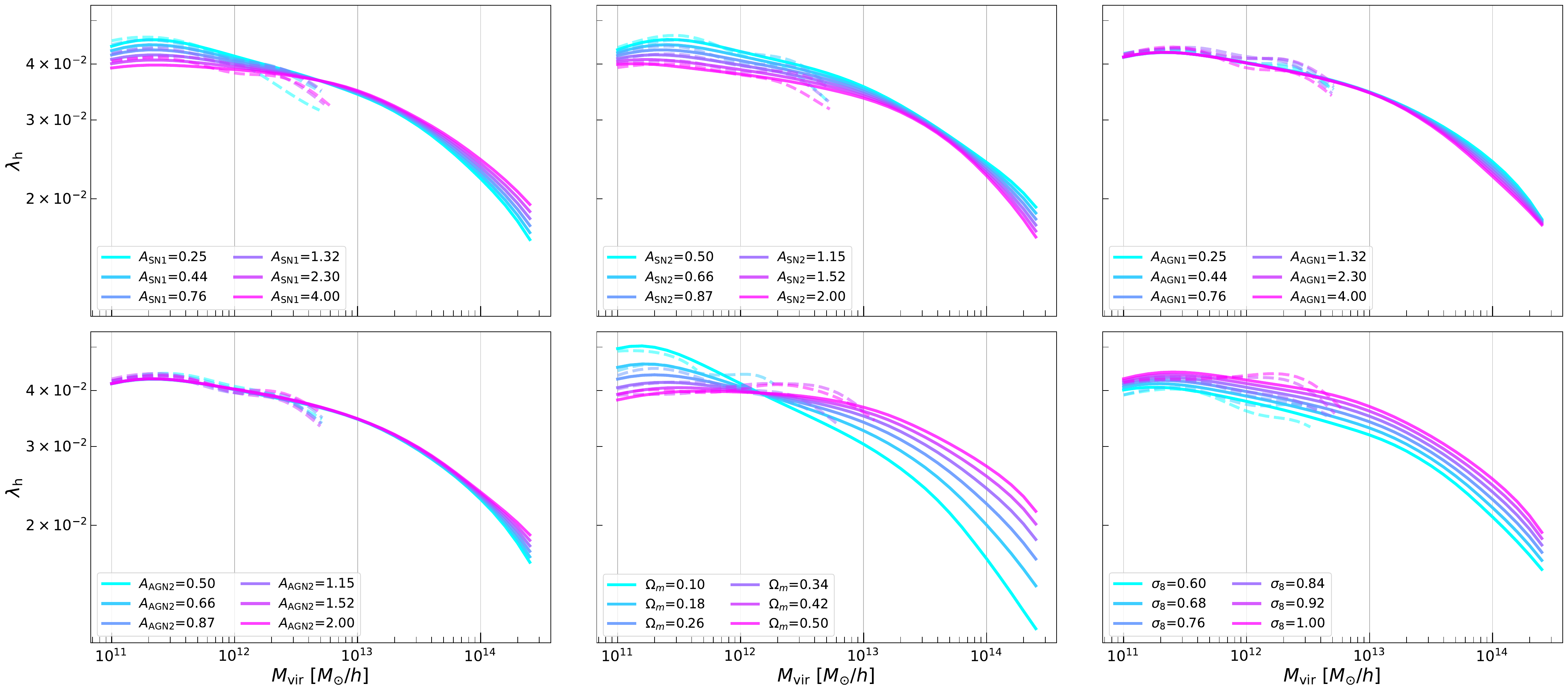}
    \caption{Same as Figure~\ref{fig:TNGsigmaDM} but for the $\spin-\Mvir$ scaling relation. The offsets between our model and the 1P scaling relations are $2-5\%$ while the $1\sigma$ from cosmic variance is $4\%$.}
    \label{fig:TNGspin}
\end{figure*}

Figure \ref{fig:TNGspin} shows the halo spin--mass relation, which has a monotonic, but nonlinear, behavior across the full mass range. Increases in $\sigma_8$ increase the $\spin$ for all masses considered, and the slope $\alpha_{\rm \sigma_8}$ is the largest amongst the six simulation parameters (see slopes in Figure \ref{fig:TNGslope}). The influence of $\Omega_{\rm m}$ has a crossover point at $\Mvir \approx 10^{12} \msol/h$; the impact is negative below this mass and positive above it. Given $\Omega_{\rm b}$ is fixed in \textsc{Camels}, the variation in $\Omega_{\rm m}$ is a variation in $1/f_{\rm b}$. Thus, variation in this parameter will also amplify/suppress any of the baryonic effects in the simulation. This could explain the existence of a crossover point, as well the mass scale of $10^{12} \msol/h$, which coincidences with the peak of the stellar mass fraction (see Figure \ref{fig:TNGMstar}) and with the onset of AGN feedback in TNG \citepalias{\Anba}. \citet{Maccio2008CosmoDep} find that spin in dark matter-only simulations has a weak dependence on cosmology and mass, and \citet{Bryan2013BaryonImpactOnShapes} also find a weak dependence on cosmology in full hydrodynamic simulations. Both tests of cosmology explore a narrower range of cosmology parameters than the range probed in our work, and a narrower range would reduce the amplitude of the measured cosmology-dependent shifts. These works find a weak dependence on halo mass, further supported by works like \citet{Knebe2008Spin}, and corresponds to slopes of the order $\alpha_M \approx -0.01$. Our work finds the spin to decrease by a factor of two over three decades in $\Mvir$, and our estimated slope (see Figure \ref{fig:TNGslope}) is of similar order for halos near MW and galaxy group scales, but is much larger at cluster scales. Note that the spin estimate provided by \textsc{Camels} includes all matter (DM, stars, and gas) whereas these prior works quoted above measure the spin using just the DM, and this difference could lead to different slopes of the $\spin - \Mvir$ relation. The computational requirements for replicating the measurements of these past works on all the \textsc{Camels} LH simulations are beyond the scope of this work. We also note that the scatter in the $\spin-\Mvir$ relation is $\approx 60\%$ which is similar in magnitude to the change we observe across the three decades in mass.

The SNe feedback parameters scale inversely with the spin at low masses: more feedback results in less coherent rotation, and reduces the spin. As shown in \citetalias{\Anba}, we expect SNe feedback to be most impactful in halos below the MW mass. For $A_{\rm SN1}$, the impact reverses sign at higher masses. This could be related to the non-linear coupling of SNe and AGN discussed in Section \ref{sec:aform}, where higher SNe feedback reduces the AGN feedback. Note also the similarity in the variation of spin with $A_{\rm SN1}$ and with $\Omega_{\rm m}$ (better thought of as $1/f_{\rm b}$ for this discussion): both have a  crossover point, though at slightly different masses. In $A_{\rm SN2}$, the inverse scaling --- where $\spin$ decreases with increases to $A_{\rm SN2}$ --- is found throughout the full mass range, with no crossover point. As we have noted before, the impact of $A_{\rm SN2}$ on the halo property is through its coupling to AGN feedback.

The AGN feedback parameters themselves have only a weak impact on the spin. There is a slight variation at high masses, and we measure the slopes with both parameters, $\alpha_{\rm AGN1/2}$, to be non-zero at $2\sigma$. At $\Mvir \approx 10^{12}\msol/h$, the AGN jet feedback begins to play a role \citepalias[][see their Figure 7]{\Anba}, and will inject kinetic energy into the gas component.

\citet[][see their Figure 3]{Bryan2013BaryonImpactOnShapes} find that the inclusion of baryons does not lead to a significant change in the $\spin-\Mvir$ relation, where $\spin$ in their work is always computed using \textit{only the DM particles}. However, their sample has a median mass of $\approx 10^{12} \msol/h$, which does not probe the $\Mvir = 10^{11}\msol/h$ halos whose spins have a stronger dependence on baryons as we find here, and their halo sample is $\mathcal{O}(10^3)$ times smaller than the one we use here, meaning the uncertainties on their estimated slopes/behaviors is larger than the trends we observe.

\subsection{Stellar mass, $\Mstar$}\label{sec:Mstar}

\begin{figure*}
    \centering
    \includegraphics[width = 2\columnwidth]{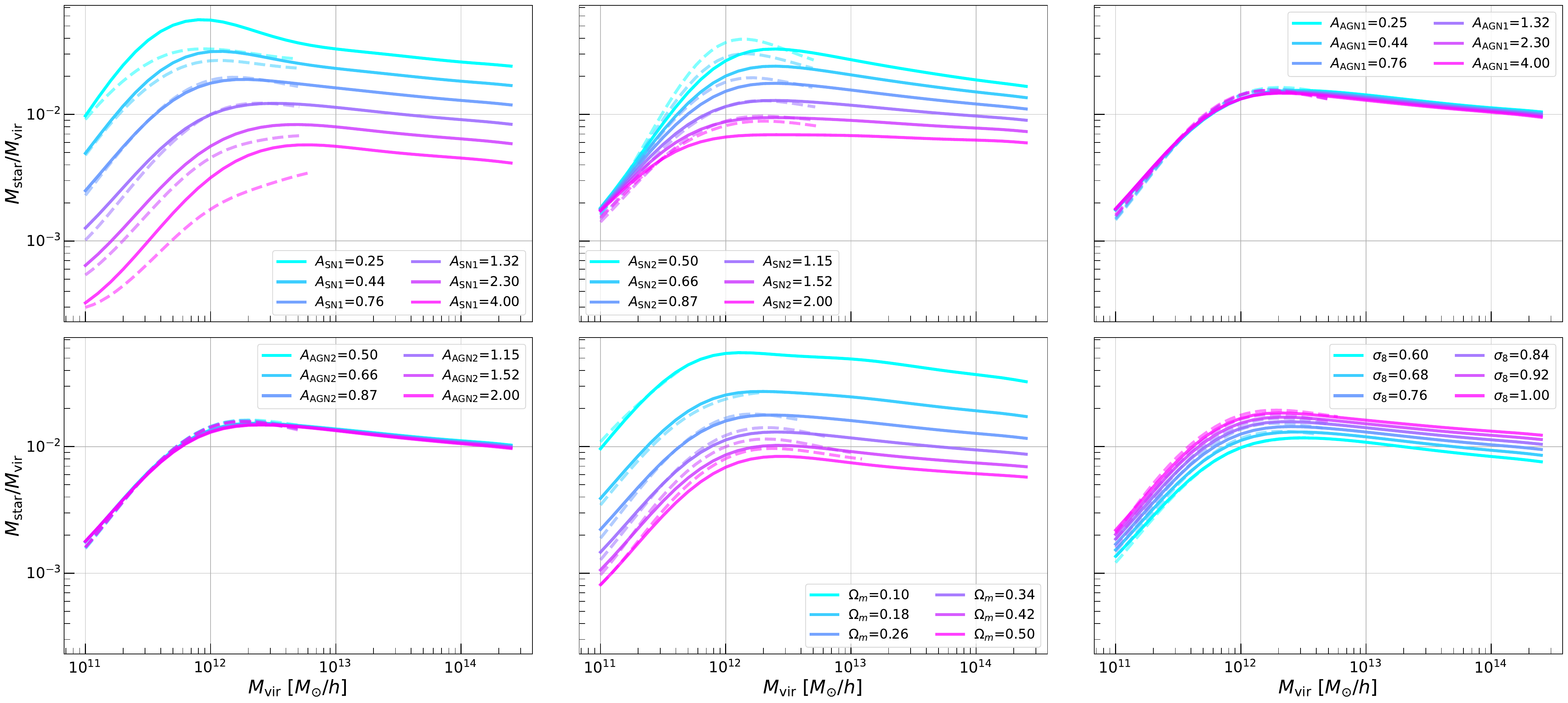}
    \caption{Same as Figure~\ref{fig:TNGsigmaDM} but for the $\Mstar-\Mvir$ scaling relation. We normalize $\Mstar$ by $\Mvir$, so the variation across mass can be more easily seen. Thus, we plot the stellar mass \textit{fraction} rather than the stellar mass. The offsets between our model and the 1P scaling relations are 5-10\% while the $1\sigma$ from cosmic variance is $5\%$}
    \label{fig:TNGMstar}
\end{figure*}

Moving to the baryonic quantities, Figure \ref{fig:TNGMstar} shows the stellar mass fraction--halo mass relation. The qualitative shape of the relation is similar under variations of all six simulation parameters: there is a steep increase at lower masses, a peak around MW mass scales, and a shallow dropoff towards higher masses. The exact mass scales of each feature vary with changes to some of the simulation parameters. The slope of $\Mstar$ (i.e. the stellar mass, not the stellar fraction) with $\Mvir$ approaches $2$ at the lowest masses and then tends toward 1 from below starting around $\Mvir = 10^{13} \msol/h$ (Figure \ref{fig:TNGslope}). The latter, asymptotic behavior is consistent with previous works that used the \textsc{Kllr} to find this trend in multiple hydrodynamical simulations \citep{Farahi2018BAHAMAS, Anbajagane2020StellarStatistics}.

The impact of $\Omega_{\rm m}$ on the stellar fraction is approximately scale-independent with a slope $\alpha_{\rm \Omega_m} \sim -1$, though it tends towards $\alpha_{\rm \Omega_m} \sim -1.5$ at the lowest masses probed here (see Figure \ref{fig:TNGslope}). Recall that in \textsc{Camels}, $\alpha_{\rm \Omega_m} = -\alpha_{\rm f_b}$ as discussed above and so $\alpha_{\rm f_b} \sim 1$. Thus, the variation with $\Omega_{\rm m}$ mentioned above can be removed through a simple rescaling of $\Mstar \rightarrow \Mstar f_{\rm b}$, and this a commonly used technique when comparing baryonic quantities across different simulations \citep[\eg][]{Lim2021GasMass, Lee2022rSZ}. The dependence on $\sigma_8$ is $\alpha_{\sigma_8} \approx 1$ for the full mass range. \citetalias{\Singh} finds that $\Mstar$ scales as $\Mstar\propto f_b^{2.5}$ and $\sigma_8^{1.7}$, and both scaling exponents are nearly twice as large as those found in our work at similar masses ($\alpha_{f_b} \approx 1$ and $\alpha_{\sigma_8} \approx 1$). The two estimates for $\alpha_{\sigma_8}$ are within $2\sigma$ of each other and thus still in the regime of statistical consistency. However, the ``cosmology dependence'' studied in baryonic simulations includes non-linear coupling between many galaxy formation processes. For example, \citet[][see their Figure 5 and B2]{\Shao} found the cosmology dependence in the \textsc{Camels} hydro simulations is significantly different from that in the associated dark matter-only simulations. Given this non-linear coupling, differences in the simulated galaxy processes can cause differences in the manifested cosmology dependence. \citet[][see their Figure 7]{Anbajagane2020StellarStatistics} found that the galaxy formation implementation in the \textsc{Magneticum} simulations, which are used in \citetalias{\Singh}, resulted in correlations between stellar properties that were sometimes qualitatively different from those of other simulations, including the \textsc{IllustrisTNG} model discussed in this work. This could lead to the apparent differences in the cosmology scaling of these baryonic properties.

The SNe feedback parameters both scale inversely with $\Mstar$ and differ in the mass range where they are relevant. The scaling with $A_{\rm SN1}$ follows the same trend as that with $\Omega_{\rm m}$, and is nearly constant above MW masses with $\alpha_{\rm \Omega_{\rm m}} \approx -0.6$ and drops to a slope of $-1.5$ below that mass-scale. The scaling with $A_{\rm SN2}$ is also roughly constant above that mass-scale, and rapidly drops to a slope, $\alpha_{A_{\rm SN2}} = 0$, below it. Increases in the SNe feedback amplitudes cause the stellar material to be more dispersed into the halo, and lead to weaker subsequent star formation. This negative feedback loop of higher $\Mstar$ leading to more SNe events is detailed in \citet{Muratov2015GasOutflowsFIRE}, where a suite of high-resolution, zoom-in simulations is used to show that gas dynamics and star formation combine into a regulatory mechanism that preserves the overall scaling of the $\Mstar-\Mvir$ relation. 

The slopes of AGN feedback parameters with $\Mstar$ are essentially zero for the full mass range. We reiterate that this only implies an insensitivity of $\Mstar$ to the \textit{specific processes} being varied here; baryonic matter is still significantly impacted by the presence of AGN feedback.

\subsection{Gas mass, $\Mgas$}\label{sec:Mgas}
\begin{figure*}
    \centering
    \includegraphics[width = 2\columnwidth]{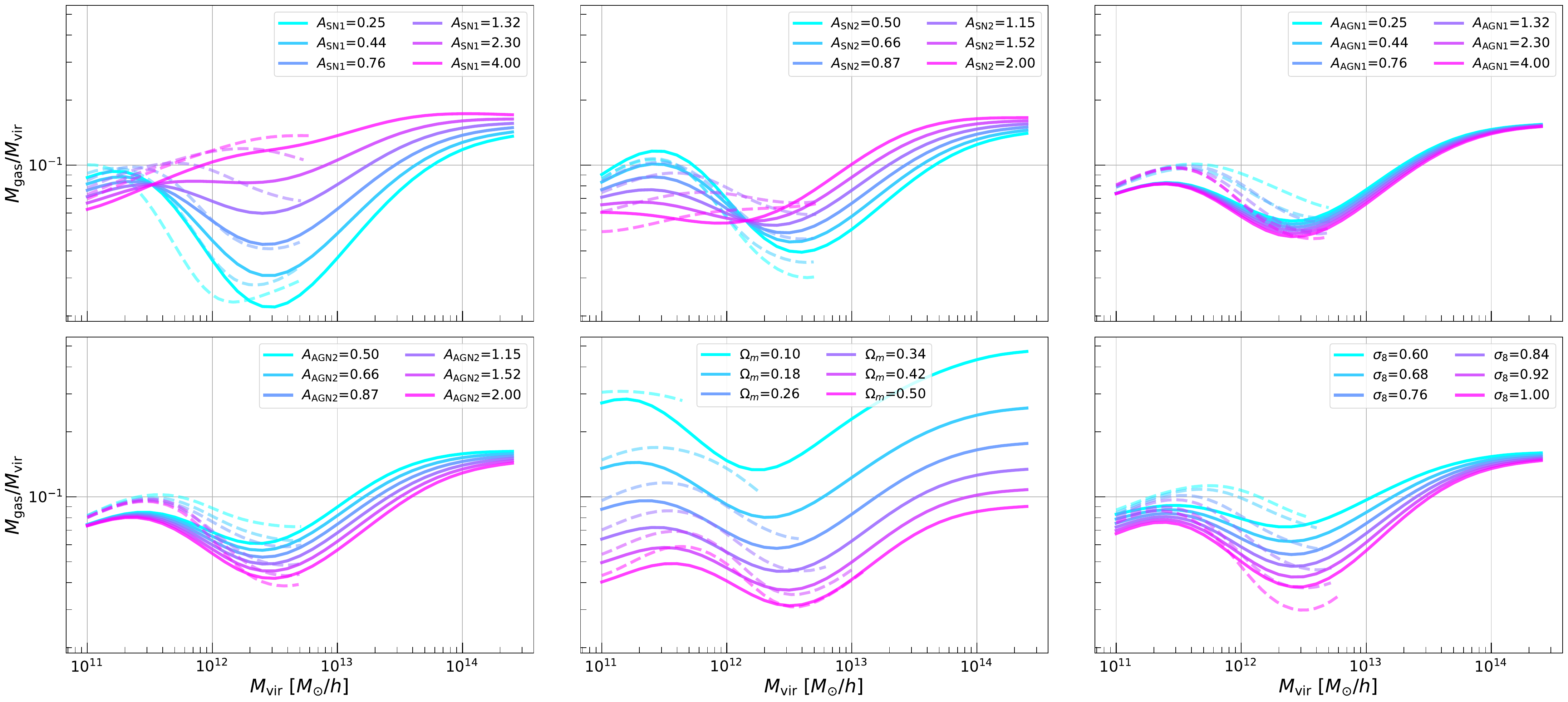}
    \caption{Same as Figure~\ref{fig:TNGsigmaDM} but for the $\Mgas-\Mvir$ scaling relation. We normalize $\Mgas$ by $\Mvir$, so the variation across mass can be more easily seen. Thus, we are actually plotting the gas mass \textit{fraction} rather than the gas mass. The offsets between our model and the 1P scaling relations are 10-20\% while the $1\sigma$ from cosmic variance is $6\%$.}
    \label{fig:TNGMgas}
\end{figure*}

\begin{figure*}
    \centering
    \includegraphics[width = 1.6\columnwidth]{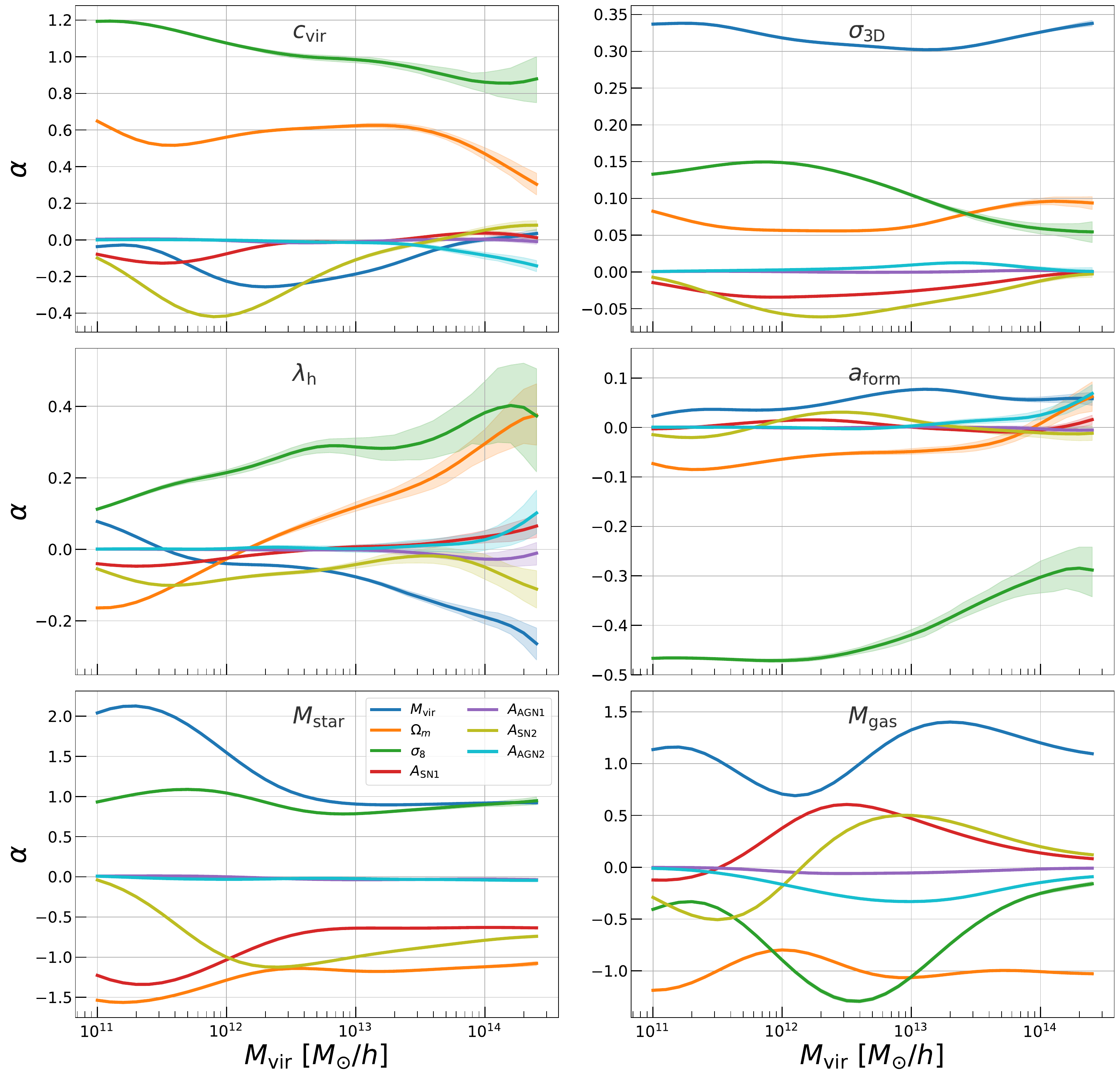}
    \caption{The \textsc{Kllr} slopes of different halo and galaxy properties with halo mass $\Mvir$ and with the 6 astrophysical and cosmological parameters in the TNG model. Each panel corresponds to one halo property as indicated by the label on top, each color represents the slope with respect to one parameter, and the bands show the 68\% confidence interval. The slopes show strong non-monotonic features for every property.}
    \label{fig:TNGslope}
\end{figure*}

Finally, in Figure \ref{fig:TNGMgas} we show the gas mass fraction--halo mass scaling relation. The slope with $\Mvir$ changes significantly over the mass range but with a median of around $\alpha = 1$ (see Figure \ref{fig:TNGslope}). At the high mass end, the slopes is $\alpha\sim 1.5$ and tends towards $\alpha \sim 1$, consistent with the results of \citet{Farahi2018BAHAMAS}. The gas fraction dips to a minimum at MW masses.

Similar to the case with $\Mstar$, we find the slope between $\Mgas$ and $\Omega_{\rm m}$ to be $\alpha_{\rm \Omega_m} \approx -1$, which can be removed through a simple rescaling of $\Mgas \rightarrow \Mgas f_{\rm b}$. The gas mass also scales inversely with $\sigma_8$, and shows a very strong mass trend, where the slope is $-0.5$ at both high and low masses and then drops to $-1.2$ near MW masses. The gas dynamics in and around the halo depend on the large-scale inflows of cold gas from the surrounding environment \citep[\eg][]{Baxter2021ShocksSZ, Aung2020SplashShock, Anbajagane2022Shocks, Anbajagane2023Shocks}, and at fixed halo mass the rate of inflow can be parameterized by $\sigma_8$. Interestingly, increasing $\sigma_8$ \textit{decreases} the gas mass fraction whereas it \textit{increases} the stellar mass fraction. \citetalias{\Singh} find $\Mgas\propto f_b^{0.8}$,  consistent with our results. They also indicate $\Mgas$ is only weakly dependent on $\sigma_8$ when considering the uncertainty on their measurement, with $\Mgas\propto \sigma_8^{-0.14}$. However, their halo sample is limited to  $\Mvir > 2\times10^{14} \msol$. Our result for this mass range is approximately $\alpha = -0.2$ (See Figure \ref{fig:TNGslope}), which is consistent with their scaling to within $1\sigma$.

The SNe parameters have strongly non-linear relations with mass. Both parameters decrease $\Mgas$ at low halo masses and increase it towards high halo masses. The former behavior is related to the ability of SNe explosions to eject gas out of the shallow halo potential of low mass objects. The latter behavior is consistent with a non-linear coupling of SNe and AGN as shown in \citet{Gebhardt2023CamelsAGNSN}, and also discussed above. Stronger AGN energetics naturally lead to a lower baryon fraction; see \citet[][see Figure 2]{Lim2021GasMass} for a comparison of baryon fractions in various simulations. SNe events also increase the available gas via the conversion of stars into disrupted material sent back out into the halo.

Finally, the two AGN parameters continue to have the weakest impact of the six simulation parameters. Though, their slope with $\Mgas$ is larger than their slopes with any other halo properties, as expected from the tight connection between AGN physics and gas dynamics. The impact of the $A_{\rm AGN2}$ is strongest at $\Mvir = 10^{13} \msol/h$ while that of  $A_{\rm AGN1}$ is close to zero. \citet[][see their Figure 1]{Mitchell2020EAGLE} show that the mass outflow rate into the interstellar medium has a maximum at $\Mvir \approx 10^{12}\msol$, whereas the rate of outlow out of the halo and into the large-scale structure is nearly mass-independent. Their work also shows that above (below) the mass-scale quoted above, the mass outflow is dominated by the AGN (SNe). This is consistent with the results and the picture presented above, where the AGN impact on the gas increases above MW masses.

\textbf{Uniqueness of the MW mass-scale}. Figure \ref{fig:TNGslope} summarizes the previous results by showing the slopes ($\alpha_X$ and $\alpha_M$, as denoted in Equation \eqref{eqn:Kllr}) for all scaling relations discussed in this work. It also shows the $\cvir-\Mvir$ slopes as analyzed in \citetalias{\Shao}. This figure highlights the importance of the MW mass-scale when understanding the impact of baryons on halo/galaxy properties. Namely, for halos of masses $\Mvir = 10^{11.5} - 10^{12.5} \msun/h$, we find key features in all properties discussed in the subsections above. At this mass scale, the SNe dependence of $\cvir$, $\sigmaDM$, $\spin$ and $\Mstar$ is maximized. This mass coincides with crossover points in the sign of the scaling of $\spin$ with $\Omega_{\rm m}$, and of $\Mgas$ and $\aform$ with SNe parameters. This mass is where the stellar mass fraction peaks, which has been observationally confirmed through many works \citep{Behroozi2010SMHMRelation, Moster2010SMHMRelation, Yang2012GalaxyHaloConnection, Wang2013SMHMRelation, Behroozi2013StarFormationHistory, Reddick2013GalaxyHaloConnection, Moster2013SFR, Birrer2014GalaxyEvolution, Lu2015SFR, Rodriguez-Puebla2017GalaxyEvolution,  Shankar2017GalaxyProperties, Kravtsov2018SMHMRelation, Behroozi2019UniverseMachine}. In the \textsc{IllustrisTNG} model, it is the mass scale where gas cooling and star formation are most efficient and it coincides with the onset of AGN feedback processes \citepalias[][see their Figure 7]{\Anba}.

\section{Baryon imprints on $\cvir$: Dependence on other halo/galaxy properties}\label{sec:selectioneffect}

The halo properties discussed above all have significant correlations with one another, and we see similarities in baryonic imprints across different properties. Thus, the selection of halos on one of the five properties above will affect not just the scaling relation of other halo properties, but also the \textit{baryonic imprints} on said properties. While the former is an effect in the mean $X -\Mvir$ relation of the selected halo sample, the latter is an effect in the change of the $X - \Mvir$ relation with changes in parameters that control galaxy formation processes. In this case, we can rewrite Equation \eqref{eqn:Kllr} as
\begin{align}\label{eqn:GeneralKllr}
    \log_{10}Y = &\,\, \pi(\Mvir, W, z) + \alpha_M(\Mvir, W, z) \log_{10}\Mvir\nonumber\\
    & + \sum_X\alpha_X(\Mvir, W, z) \log_{10}X,
\end{align}
where the intercept $\pi$ and the slopes $\alpha$ are now all functions of the halo properties $W \in \{\cvir, \sigmaDM, \aform, \spin, \Mstar, \Mgas\}$ as well and where the set $W$ does not contain the property $Y$ that is being fit. We have thus far assumed the slopes $\alpha_X$, which define the dependence of property $Y$ on cosmology and galaxy formation parameters, do not depend on the halo properties other than $\Mvir$. However, these slopes are generically expected to depend on other properties. For example, the baryon fraction determines the specifics of different processes within the halo \citep[\eg][]{vanDaalen2020fb, Pandey2023Baryons, Grandis2023Baryons}.

The \textsc{Kllr} model defined by Equation \eqref{eqn:GeneralKllr} is computationally expensive to build, as the slopes must be computed for a grid of dimension N + 1, where N is the number of halo properties it is a function of (the extra dimension is redshift). In Equation \eqref{eqn:Kllr} this was just a 2D grid spanned by redshift and mass. In this work, we do not calibrate the full N + 1 dimensional grid and instead take the simpler approach of computing the slopes in 2D grids but for pairs of different halo samples,
\begin{align}\label{eqn:PairedKllr}
    \log_{10}Y^{W\pm} = &\,\, \pi^{W\pm}(\Mvir, z) + \alpha_M^{W\pm}(\Mvir, z) \log_{10}\Mvir\nonumber\\
    & + \sum_X\alpha^{W\pm}_X(\Mvir, z) \log_{10}X,
\end{align}
where the superscript ${W\pm}$ denotes a model corresponding to (and built with) halos where the value of a single property $W$ is higher or lower than some chosen value. We continue to use the same set of input parameters, $X \in \{\Omega_{\rm m}, \sigma_8, A_{\rm SN1}, A_{\rm AGN1}, A_{\rm SN2}, A_{\rm AGN2}\}$.

In this section, we extract the relation $\cvir(\Mvir, z, \Omega_{\rm m}, \sigma_8, A_{\rm SN1}, A_{\rm AGN1}, A_{\rm SN2}, A_{\rm AGN2})$ for halos selected on the five properties, split into selections on halo properties (which include DM and baryons) in \S\ref{sec:haloselection} and on galaxy properties (which include only baryons) in \S\ref{sec:galaxyselection}. We compute $\alpha^{W\pm}$ and discuss the changes in the estimated slopes between pairs of halo subsamples. The subsample pair is constructed by taking the uber sample of all halos from the 1000 LH simulation. Then, for each halo and chosen selection property, $W$, we predict the mean property value given the halo mass, redshift, and six input simulation parameters. The models used to predict the mean are the same ones shown above in Section \ref{sec:AstroCosmoDependence}. Each halo is then placed into one of two subsamples (denoted by $W\pm$) depending on whether the value of $W$ for that halo is above/below the predicted mean. We then rebuild the $\cvir-\Mvir$ model separately for each subset, which involves estimating the slopes $\alpha^{W\pm}$ in Equation \eqref{eqn:PairedKllr}.

Note that such a split will cause a difference in the mean $\cvir-\Mvir$ relation, i.e. the intercept $\pi^{W+}(\Mvir, z)$ will differ from $\pi^{W-}(\Mvir, z)$, in a way defined by the correlation between the property $Y$ and the split variable $W$. The impact of this for correlated variables is detailed in the formalism of \citet{Evrard2014MultiProperty}. While we will briefly discuss the impact of selection effects on $\pi^{W\pm}(\Mvir)$ in Figure \ref{fig:MeanCvirSplit}, our discussions focus on $\alpha^{W\pm}(\Mvir)$ as this quantifies changes in the baryonic imprints due to different halo/galaxy selections.

\begin{figure*}
    \centering
    \includegraphics[width = 2\columnwidth]{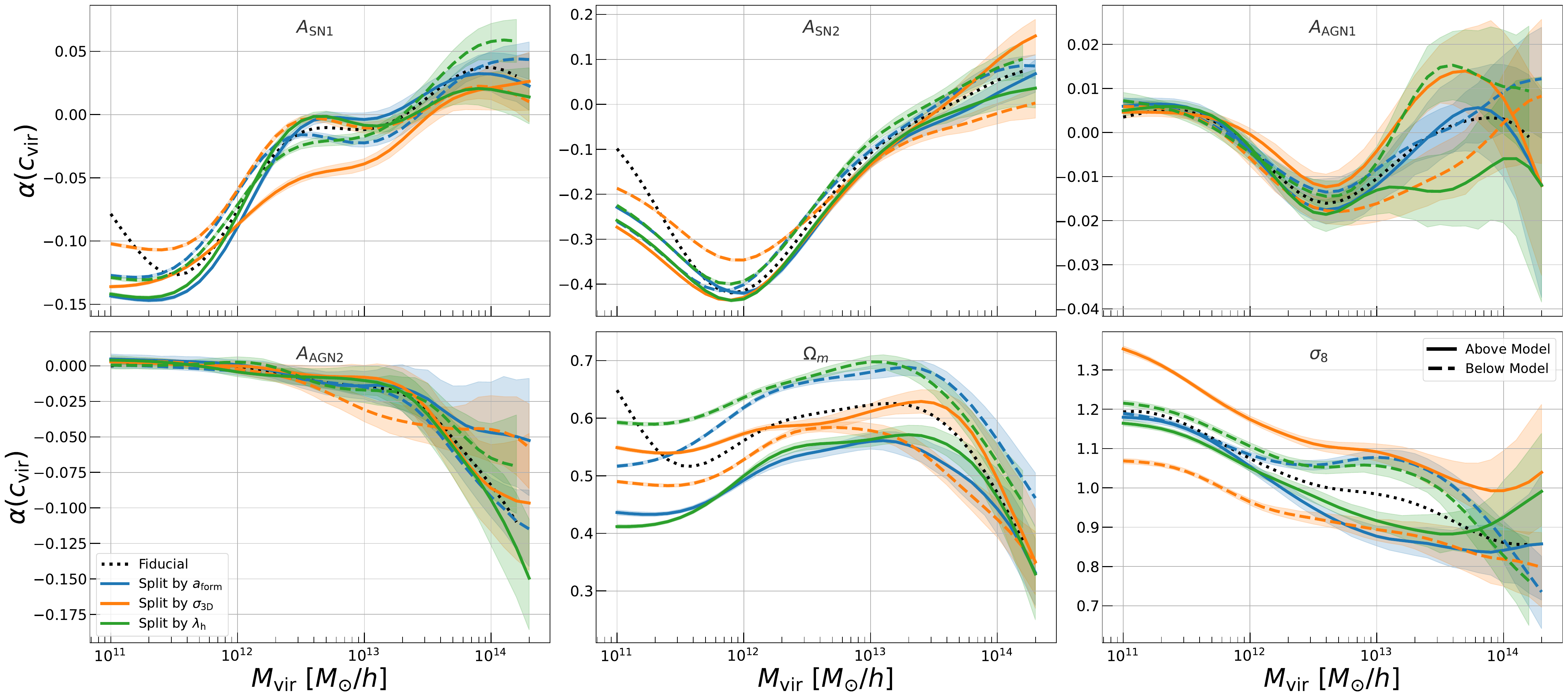}
    \caption{The slopes of $\cvir$ with the six parameters, for halo samples split by a given property (colored lines), where the split is on halos that lie above/below the mean relation of a given property. The fiducial slope, estimated using the full, unsplit halo sample, is shown in the black dotted line. The scaling with astrophysics parameters is mostly unchanged through this selection, while that with cosmology parameters is modified significantly. All model predictions were obtained using the standard values of the input parameters (see Section \ref{sec:Camels}).}
    \label{fig:split_halo}
\end{figure*}

\subsection{Halo properties}\label{sec:haloselection}

\begin{figure*}[!ht]
    \centering
    \includegraphics[width = 2\columnwidth]{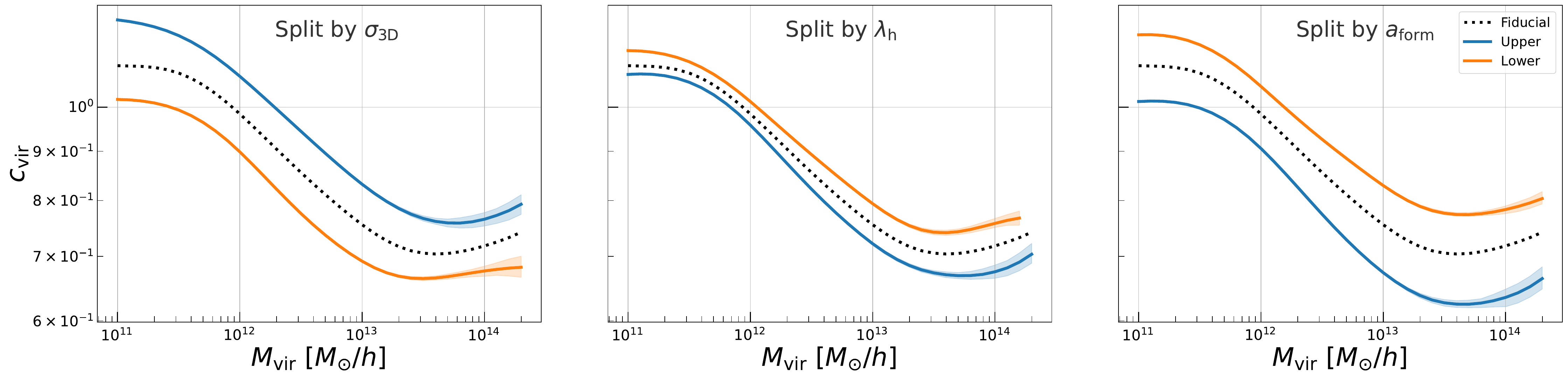}\\[10pt]
    \includegraphics[width = 1.35\columnwidth]{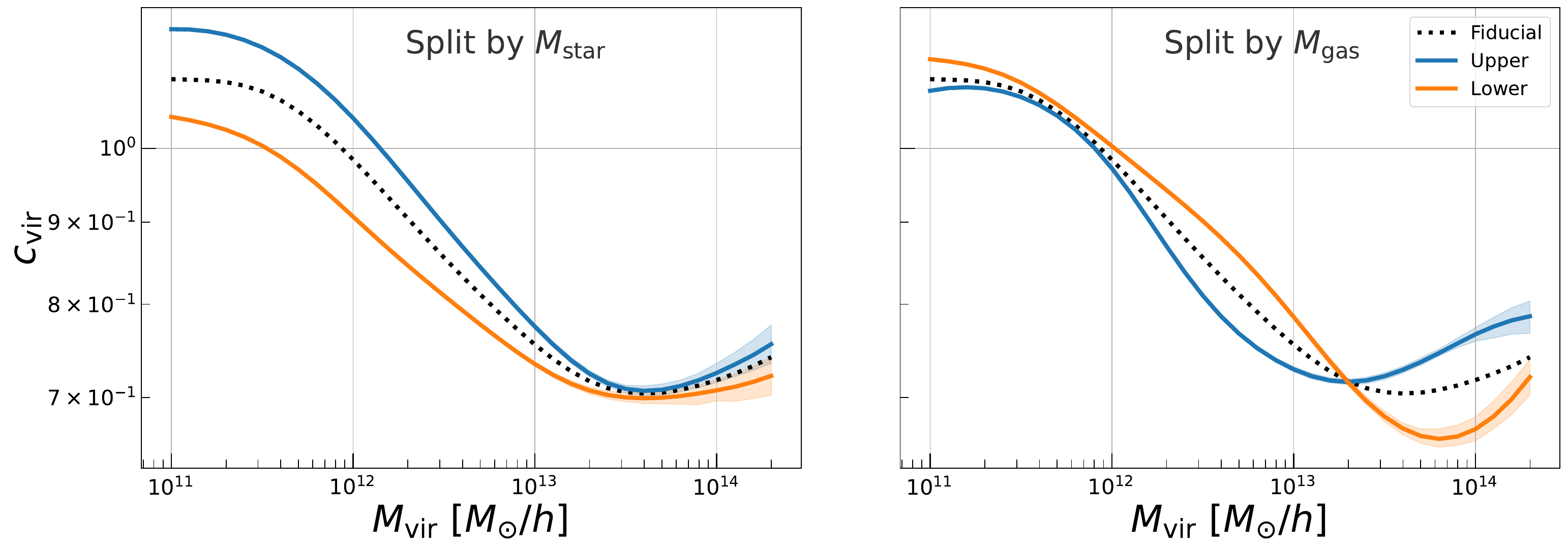}
    \caption{The $\cvir-\Mvir$ relation for two halo subsamples, defined as halos with a property value above/below the mean value at that mass. All splits show clear differences in at least some part of the halo mass range. The split on $\Mgas$ causes a crossover in $\cvir$ towards higher halo masses. All model predictions were obtained using the standard values of the input parameters (see Section \ref{sec:Camels}).}
    \label{fig:MeanCvirSplit}
\end{figure*}

Figure \ref{fig:split_halo} presents the slopes of the halo concentration with respect to the four astrophysical and two cosmological parameters. The slopes are shown for halos split on different properties, as defined in Equation \eqref{eqn:PairedKllr}. The changes in mass dependence are shown in Figure \ref{fig:MeanCvirSplit}. In Figure \ref{fig:split_halo}, the solid (dashed) curves correspond to the subset of halos with a specific property value greater (lesser) than the mean. The average of the two slopes is not guaranteed to match the slope of the fiducial sample. The mean relations can follow such a behavior but the slopes, which are the derivative of the mean relation with respect to mass and thus an estimate of the \textit{shape} of the relation, need not follow such behavior.

Focusing first on the cosmology dependence, selecting halos on $\aform$ and $\spin$ has a significant impact on $\alpha_{\Omega_{\rm m}}$ while $\aform$, $\spin$, and $\sigmaDM$ all affect $\alpha_{\sigma_8}$. The difference in $\alpha_{\Omega_{\rm m}}$ is nearly mass-independent. The results for $\alpha_{\sigma_8}$, however, show a strong mass-dependence when split on $\spin$ and $\aform$, while the split on $\sigmaDM$ still leaves a nearly mass-independent effect. All three of $\aform$, $\spin$, and $\sigmaDM$ are highly correlated with the mass accretion rate \citep{Lau2020HaloShapesCorrelations, Anbajagane2022BaryImprint}, as halos with a high mass accretion form earlier (lower $\aform$) and have more concentrated potentials \citep{Wechsler2002Concentrations}, leading to higher velocity dispersions \citep[\eg][]{Okoli2016Concentration}. Halos with a high accretion rate have been shown to have a high spin \citep[][see their Figure 3]{Johnson2019SpinMAR}. Thus, all three selections above are correlated with a property of the large-scale environment of the halo, namely the halo mass accretion rate. This suggests that the relation between $\cvir$ and cosmology could be altered depending on the accretion state (or large-scale environment) of the halo. We leave a detailed analysis of the causal origin of this correlation to future work and we will simply employ the empirical results in Figure \ref{fig:split_halo} to test the impact of selection effects further below.

Moving onto the SNe feedback, we see that most selections cause very little change as the slopes of the subsamples overlap with that of the fiducial relation (black dotted line). All halo samples show the same phenomenological features, such as local minima and maxima. The main differences occur from selecting halos on $\sigmaDM$. We see that in high $\sigmaDM$ systems, the concentration is less dependent on $A_{\rm SN1}$ for $10^{12} \msol/h < \Mvir < 10^{13}\msol/h$, and in low $\sigmaDM$ systems it is less dependent on $A_{\rm SN2}$ for $\Mvir < 10^{12.5} \msol/h$. The SNe wind speed modeled in \textsc{IllustrisTNG} relies on the local (dark matter) velocity dispersion around the stars \citep[][see their Equation 1]{Pillepich2018Methods}. This implementation is based on previous semi-analytic models that found such a scaling was needed to reproduce the observed stellar mass functions and restframe B-band and K-band luminosity functions across redshift \citep{Henriques2013SAM}. Given this choice, it is natural for the scaling with $A_{\rm SN2}$, where this parameter controls the wind speed amplitude \citep[][see their Equation 4]{Villaescusa-Navarro2021CAMELS}, to depend $\sigmaDM$ selections.

Finally, focusing on AGN feedback, the halo selections cause no statistically significant difference in the scaling of $\cvir$ with the AGN-related parameters. The strongest feature (which is still only a $2\sigma$ difference) is that halos of high $\sigmaDM$ or low $\spin$ have $\cvir$ increase with $A_{\rm AGN}$.

In Figure \ref{fig:MeanCvirSplit}, we also compare the mean $\cvir-\Mvir$ scaling relations of the halo samples split on $\aform$, $\sigmaDM$ or $\spin$. These splits result in clear differences in the $\cvir-\Mvir$ relations that are nearly constant across the whole mass range. The shape of the relation is similar across all three subsamples, meaning the slope/shape of $\cvir-\Mvir$ does not change much with selections on these three properties. Similar to our discussion above, we find $\cvir$ is positively correlated with $\sigmaDM$ and negatively correlated with $\aform$ and $\spin$, consistent with previous work \citep[\eg][]{Wechsler2002Concentrations, Lau2020HaloShapesCorrelations, Hearin2021DiffMAH, Anbajagane2022BaryImprint}.

\subsection{Galaxy properties}\label{sec:galaxyselection}

Finally, we consider the impact of baryonic property selections on the slopes $\alpha_X$. Under this selection, all slopes experience significant changes as seen by the differences between the dashed and solid lines in each plot; as expected given the direct connection between galaxy formation processes and baryonic properties.

\begin{figure*}
    \centering
    \includegraphics[width = 2\columnwidth]{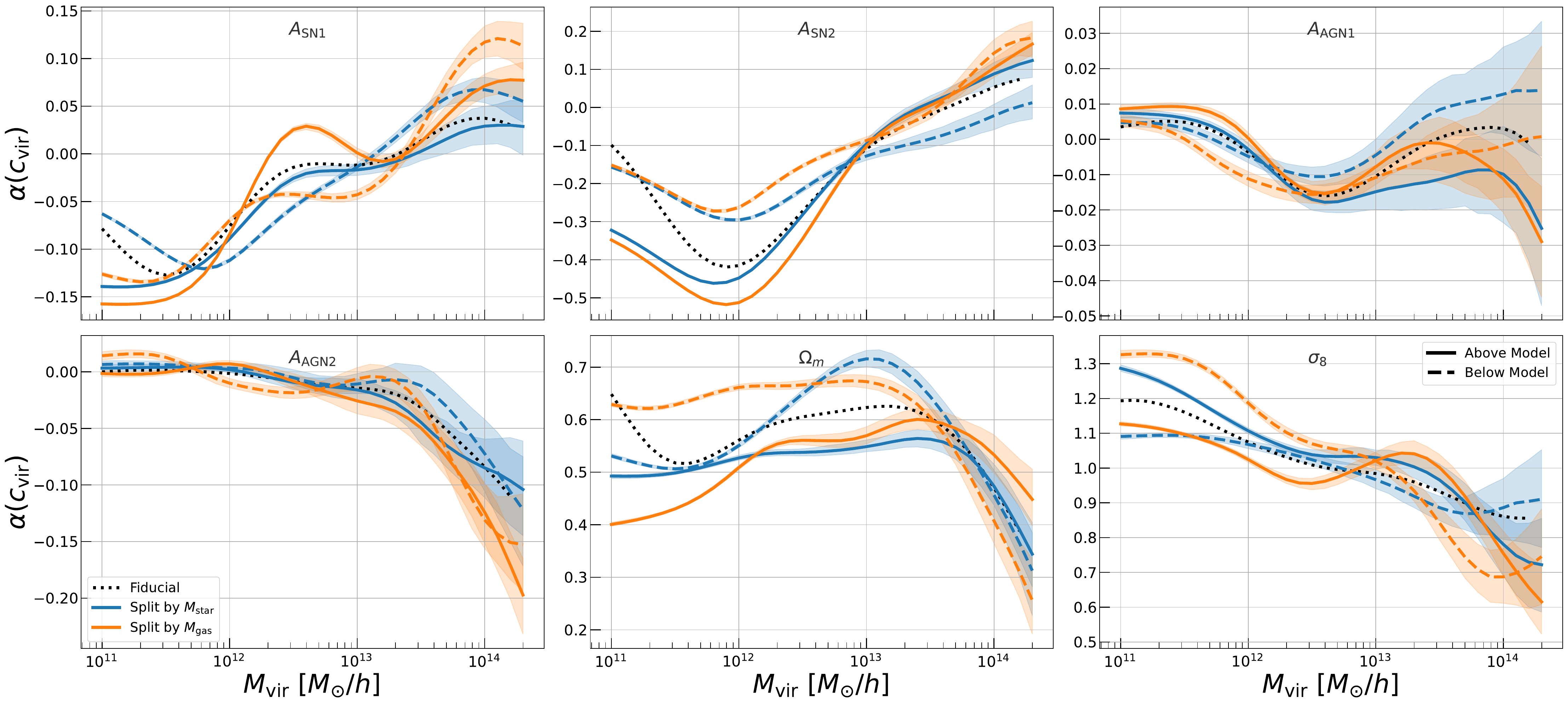}
    \caption{\centering Similar to Figure \ref{fig:split_halo} but for selections/splits on $\Mstar$ and $\Mgas$. }
    \label{fig:split_galaxy}
\end{figure*}

There are significant differences in the scaling of $\cvir$ with cosmology parameters when comparing the two subsamples. Starting with $\Omega_{\rm m}$, we find that the $\Mgas$ selection is most impactful, particularly for $\Mvir < 10^{13.5} \msol/h$. The $\Mstar$ selection mimics the $\Mgas$ one in the range $10^{12} \msol/h < \Mvir < 10^{13.5} \msol/h$, suggesting that halo baryon selection is the key factor here (since we find halos with high $\Mstar$ or high $\Mgas$ to cause similar changes in the slopes). The change in $\alpha_{\sigma_8}$ is fairly similar across most of the mass range. Below $\Mvir < 10^{12.5} \msol/h$, we see that halos with high $\Mstar$ are qualitatively similar to those with low $\Mgas$, or vice-versa, as they both have a stronger dependence on $\sigma_8$. This sign flip in the behavior is consistent with Figure \ref{fig:TNGMstar} and \ref{fig:TNGMgas}, where we show that increasing $\sigma_8$ leads to increases in $\Mstar$ but decreases in $\Mgas$.

Moving onto the changes in the scaling between $\cvir$ and SNe parameters, we find some variations due to selections on stellar/gas mass and they are all statistically significant. For $A_{\rm SN1}$, though, the changes are small in amplitude (the slope varies by $\Delta\alpha < 0.05$). The $A_{\rm SN2}$ dependence is much more significant, where halos with low stellar/gas mass have a lower dependence on this parameter. Since SNe winds depend on the amount of stars, and the impact of baryons on the concentration is through redistribution of the gas, reducing $\Mstar$ or $\Mgas$ reduces the impact of this parameter on $\cvir$.

The AGN parameters once again have the smallest variations. The gas selection impacts the $A_{\rm AGN1}$ and $A_{\rm AGN2}$ dependence at low masses, $\Mvir < 10^{12}\msol/h$. We do not discuss this further given the low amplitudes of these parameters.

In Figure \ref{fig:MeanCvirSplit}, we also compare the $\cvir-\Mvir$ scaling relations of the halo samples split on $\Mstar$ and $\Mgas$. These splits result in clear differences in the $\cvir-\Mvir$ relation, and differences that have unique a mass-dependence. Splitting on $\Mstar$ induces a larger change in the $\cvir$ normalization for $\Mvir < 10^{13} \msol/h$, and almost no significant change above that mass. Splitting on $\Mgas$ causes smaller changes at $\Mvir < 10^{12} \msol/h$ and large changes above this mass scale. This latter split also causes a crossover at $\Mvir \approx 10^{13.5}\msol/h$. Above (below) this mass, halos with more gas have a higher (lower) concentration. In the TNG model, the internal energetics of halos above (below) this mass-scale are dominated by gas cooling (AGN feedback); see Figure 7 in \citetalias{\Anba}. When selecting halos with more gas at fixed halo mass, we are selecting halos whose gas fraction is higher, which means the gas can affect the internal dynamics more. At high (low) masses the dominance of gas cooling (AGN feedback) means more of the matter can be concentrated in the halo core (dispersed out of the halo core), which leads to an increase (decrease) in $\cvir$. The same behavior does not occur for the $\Mstar$ split as stellar matter is both a much smaller fraction of the total matter, and does not directly couple to feedback.

\textbf{Estimate of potential selection-based modelling error.} As expected, selections on gas mass or stellar mass result in different slopes, $\alpha_X$, between $\cvir$ and the galaxy formation parameters $X$. An accurate treatment of baryon imprints on the density field around halos, where the halos are selected by an observable connected to emission from the gas or stellar component, would require a more general model than a mass-dependent one alone, i.e. the slopes, $\alpha_X$, cannot be represented as just functions of halo mass. We make a quantitative estimate of this by computing the differences between two predictions for various parameters. This can be written as
\begin{align}\label{eqn:diff}
    \Delta  \log_{10} \cvir (X, W) = & \,\, (\alpha_M^{W+} - \alpha_M^{W-})\log_{10}\Mvir \nonumber\\
    & + \sum_X(\alpha_X^{W+} - \alpha_X^{W-})\log_{10}X
\end{align}
where $X$ is a set of values for the six simulation parameters, and $W$ is the property we perform the split with. The individual slopes, $\alpha^{W\pm}$, are shown in Figure \ref{fig:split_halo} and \ref{fig:split_galaxy}. By computing $\Delta \log_{10} \cvir$ for various choices of input $X$, we can quantify the possible modeling errors caused by the selection, $W^\pm$. These can be thought of as errors arising from an improperly specified model \eg the theoretical error from a model built on halos with high $W$ being applied to halos with low $W$. Equation \eqref{eqn:diff} only includes the slopes and not the normalization, $\pi^{W\pm}$, as only the former captures the response of the $\cvir-\Mvir$ relation to changes in galaxy formation and cosmology. We have also already shown in Figure \ref{fig:MeanCvirSplit} the selection-based differences in the normalization, $\pi$. Thus, the error defined in Equation  \eqref{eqn:diff} is for models that specify how the $\cvir-\Mvir$ relation responds to changes in the six simulation parameters.

\begin{figure}
    \centering
    \includegraphics[width = \columnwidth]{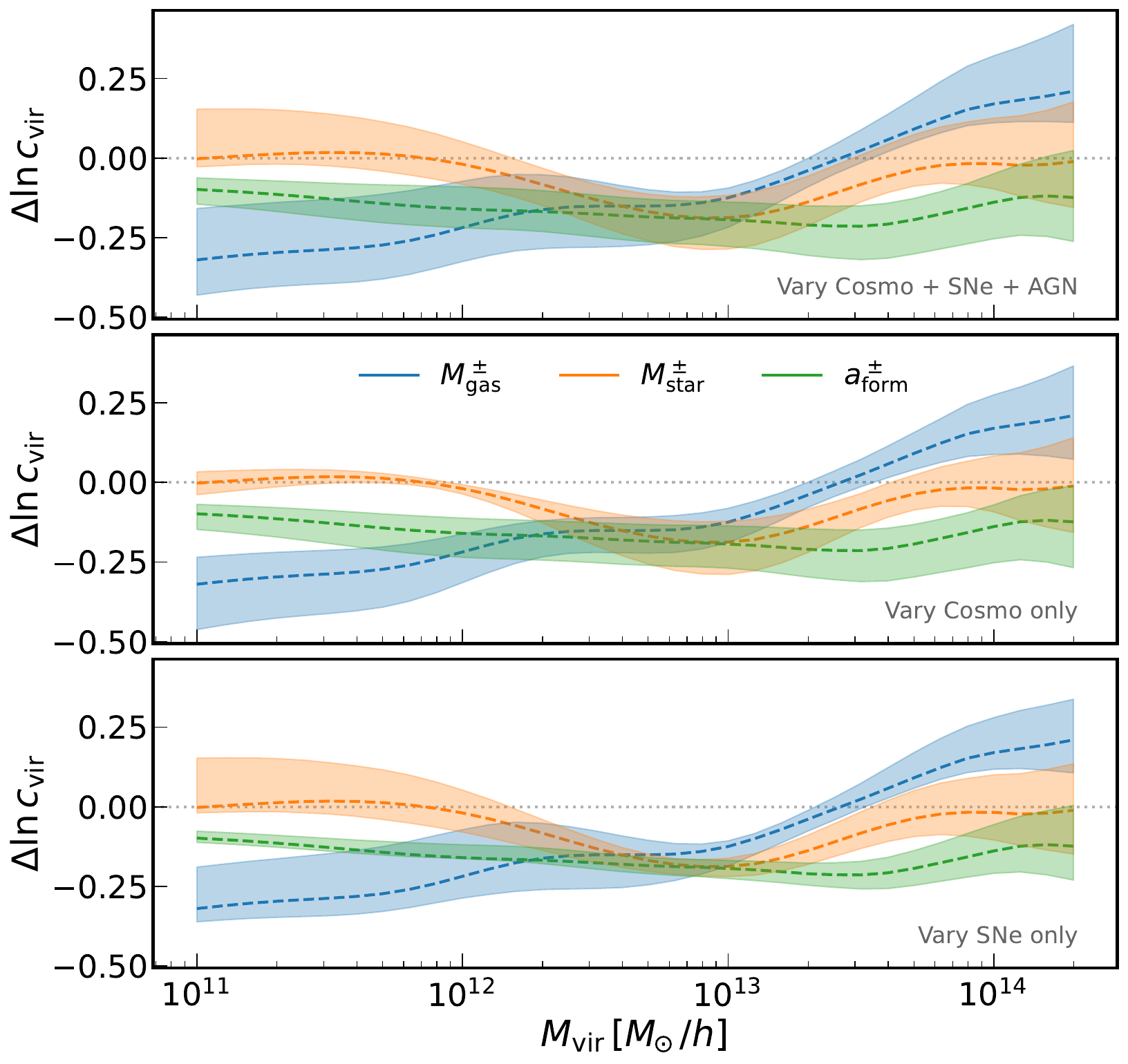}
    \caption{The fractional modeling error, $\Delta \ln \cvir$, due to ignoring selection effects on a given halo property. If the value is positive (negative), then the fiducial baryon imprints model, which accounts for no selection effects, would underpredict (overpredict) the $\cvir$ of the property-selected halo sample. The bands show the 68\% distribution of errors across a wide prior in the simulation parameters. The dashed lines are the error for the fiducial simulation parameter values. The top panel shows results from varying all six parameters, middle panel from varying cosmology alone, and bottom panel from varying SNe parameters alone. Selection on $\Mgas$ causes differences of the order 25\% at high and low masses. Selection on $\aform$ or $\Mstar$ (which impacts galaxy colors and luminosity) has an effect of a similar magnitude but at galaxy group scales.}
    \label{fig:Deltacvir}
\end{figure}

Figure \ref{fig:Deltacvir} presents the modelling errors in the baryon imprints due to  halo selection effects. We show $\Delta\ln\cvir = \ln(10)\Delta \log_{10}\cvir$, as it can be interpreted as a fractional difference in $\cvir$. These differences depend on the choice of input parameters and so we first sample $10^4$ points in the 6D parameter space, which is bound by $0.1 < \Omega_{\rm m} < 0.5$, $0.6 <  \sigma_8 < 1.0$, $0.25 < A_{\rm SN1},A_{\rm AGN1} < 4$, and $0.5 < A_{\rm SN2},A_{\rm AGN2} < 2$. This bound is simply the range of parameters sampled in the LH suite of the \textsc{Camels} simulations (see Section \ref{sec:Camels}). The $10^4$ points are sampled from a uniform distribution in the linear (not logarithmic) parameters. For each point, we compute $\Delta\ln\cvir$, the fractional  modelling error  in $\cvir$, as denoted in Equation \ref{eqn:diff}. Figure \ref{fig:Deltacvir} shows, as colored bands, the 68\% bounds of this error across the $10^4$ points in parameter space. The dashed lines are $\Delta\ln\cvir$ for the fiducial parameter values of  $\Omega_{\rm m} = 0.3$, $\sigma_8 = 0.8$ and $A_X = 1$.

The bands in the top panel show the range of errors in a 6D parameter space comprising all six simulation parameters, the middle and bottom panels show the same in 2D spaces varying either only cosmology parameters or only SNe parameters, respectively.  In all panels, the dashed lines show the error when assuming the fiducial simulation parameters, and the bands show the range of possible errors if we vary certain sets of parameters. The modeling errors are found to be $\approx 25\%$ in amplitude under selections on $\Mgas$ or $\aform$. This is generally lower than, but of the same order as, the scatter in $\cvir$ (which is $\approx 40\%$) and it is also comparable to differences between the $\cvir-\Mvir$ relation in dark-matter only simulations and full hydrodynamic simulations from \textsc{IllustrisTNG} \citepalias{\Anba}. Thus, selection-based errors in the baryon imprints model are not guaranteed to be subdominant, though their exact importance will depend on the dataset being modeled. The above suggests that halo selection may be a relevant factor to consider when building simulation-calibrated models for the impact of galaxy formation on the density field.

\section{Conclusion}\label{sec:conclusion}

The concentration--mass relation has ubiquitous use in modeling the matter fields in our Universe on quasi-linear and non-linear scales \citep[\eg][]{Pandey2021DESxACT, Gatti2021DESxACT, Zacharegkas2021GGLensingDES, Miyatake2023HSCY3HaloModel}. The impact of baryons on the concentration, and the matter distribution, is an active area of study \citep[\eg][]{Duffy2010BaryonDmProfileDensity, Ragagnin2019HaloConcentration, Beltz-Mohrmann2021BaryonImpactTNG, Anbajagane2022BaryImprint, Shao2023Baryons}. An equally pertinent question is the impact of baryons around \textit{certain types of halos}, such as those selected by a specific kind of observational technique --- X-ray, optical, Sunyaev Zeldovich decrement \citep{Sunyaev1972SZEffect} etc. In this work, we provide a first attempt at quantifying the impact of halo/galaxy selection on the scaling of $\cvir$ with galaxy formation parameters. We use 1000 simulations from the \textsc{Camels} suite to build a model of halo concentration ($\cvir$), velocity dispersion ($\sigmaDM$), spin ($\spin$), formation time ($\aform$), stellar mass ($\Mstar$), and gas mass ($\Mgas$). We first analyze the dependence of each individual property on galaxy formation parameters, and then also estimate the change in the slopes $\alpha_X$ --- which quantify how $\cvir$ scales with these parameters --- after selecting halos on one property. Our results are as follows:

\begin{itemize}
    \item The relationship between the halo properties and the simulation parameters is captured well through a simple local multi-linear regression. The locality of the regression is necessary as the trends are always nonlinear in halo mass and often non-monotonic as well (Figures \ref{fig:TNGsigmaDM}-\ref{fig:TNGMgas}).

    \item All halo properties have the shape of the mass-dependence altered via changes to the simulation parameters. In general, changes to the SNe feedback parameters $A_{\rm SN1/2}$ and the cosmology parameters $\Omega_{\rm m}$ and $\sigma_8$ have a significant impact on scaling relations. The AGN feedback parameters have a negligible impact.

    \item We find $\Mvir \approx 10^{12} \msol/h$ to be a key mass scale, as it is associated with (i) the strongest dependence of $\cvir$, $\sigmaDM$, $\Mstar$ on $A_{\rm SN2}$, and of $\sigmaDM$ on $\sigma_8$, (ii) a sign-flip in the dependence of $\spin$ on $\Omega_{\rm m}$, (iii) the peak of the stellar mass fraction, (iv) a sign flip in the dependence of $\aform$ and $\Mgas$ on $A_{\rm SN1}$ and $A_{\rm SN2}$. 

    \item The $\aform -\Mvir$ and $\Mgas -\Mvir$ scaling relation show a highly non-monotonic dependence on $A_{\rm SN1}$ and $A_{\rm SN2}$, and that increasing SNe feedback \textit{increases} the baryon fraction (Figure \ref{fig:TNGMgas}). This is consistent with a nonlinear coupling of SNe and AGN, as presented in \citet{Gebhardt2023CamelsAGNSN}, where the SNe takes gas from the vicinity of the AGN and sends it further out in the halo (though not out of the halo), which then prevents the AGN from ejecting material out of the halo (Figure \ref{fig:TNGaform}, \ref{fig:TNGMgas}).

    \item The baryon imprints in $\cvir$ --- captured by the scaling of $\cvir$ with galaxy formation parameters --- depend only loosely on any selections in the halo properties $\aform$, $\spin$, and $\sigmaDM$ (Figure \ref{fig:split_halo}). Instead, these selections have a large impact on the cosmology-dependence of the $\cvir-\Mvir$ relation.

    \item These imprints are more sensitive to selections on $\Mgas$ or $\Mstar$ (Figure \ref{fig:split_galaxy}). In general, halo samples with higher $\Mstar$ show a weaker dependence of $\cvir$ with $\Omega_{\rm m}$ and $\sigma_8$ and stronger dependence with $A_{\rm SN2}$. Samples of higher $\Mgas$ follow the same trend except for $\sigma_8$, where the \textit{higher $\Mgas$} sample has a weaker dependence on $\sigma_8$.

    \item The mean $\cvir - \Mvir$ relation of the split subsamples generally have the same shape but with slightly higher/lower amplitudes (Figure \ref{fig:MeanCvirSplit}). Selections on higher $\sigmaDM$ and $\Mstar$ results in higher $\cvir$, since both properties are connected to the core of the halo potential. Selecting high $\aform$ or $\spin$ results in \textit{lower} $\cvir$, while selecting high $\Mgas$ leads to lower $\cvir$ at MW masses and high $\cvir$ at cluster masses.

    \item If we ignore the impact of selection effects, the subsequent error in modeling of baryon imprints in the $\cvir - \Mvir$ relation is of the order $25\%$ for group-scale and cluster-scale halos (Figure \ref{fig:Deltacvir}).
\end{itemize}

Our results show that the baryon imprints on $\cvir$ --- defined in this work through the slopes, $\alpha_X$, which parameterize the scaling of $\cvir$ with the galaxy formation parameters --- have a clear dependence on the selection function of the sample. The most impactful selections include those on $\Mgas$ and $\Mstar$, which are also the most observationally relevant as they correspond to selections on X-ray luminosity and Sunyaev Zeldovich decrement \citep{Sunyaev1972SZEffect}, or on the galaxy brightness/magnitude, respectively. The importance of this dependence of baryon imprints on halo selection will vary from application to application, but not accounting for this selection-based effect in the baryon modeling can lead to biases. However, the exact bias depends on the range of masses and physical scales being probed in the analysis of interest.

We also build on previous works to show evidence of the non-linear coupling between different feedback mechanisms. Thus, SNe feedback may be naively thought to have no impact on the ejection of material out of the halo, whereas in practice it has a significant impact. Such couplings can be useful in the building and implementation of any semi-analytic models of galaxy formation, where AGN feedback is the dominant modelling component. We have only explored six simulation parameters (two cosmological parameters and four galaxy formation parameters). Exploring the scaling with other processes/parameters will uncover more non-linear couplings between different galaxy formation processes, and shed light on the range of processes that can alter the impact of AGN feedback in the halo's internal dynamics. Suites such as the 28-parameter variation in \textsc{Camels} will enable such analyses \citep{Ni2023Camels28}.

The improvements in computational resources have enabled the production of large simulation suites to study various physical phenomena, of both astrophysical and cosmological nature \citep[\eg][]{Navarro2020Quijote, Villaescusa-Navarro2021CAMELS, Kacprzak2023Cosmogrid, Anbajagane2023Inflation}. This plethora of data has moved forward our ability to construct data-driven, non-parametric models that capture the complexities of the scaling relations. Summarizing an aspect of these complexities, such as the non-linear mass dependence, can also help build phenomenological models that can be utilized in analyses of observational data. Our ability to build data-driven models of halo properties will grow exponentially with the expected influx of both simulated and observational data, and so studies of the robust construction and application of such models are paramount to the efficient use of simulations in constraining astrophysics and cosmology.

\section*{Acknowledgements}

We thank Andrew Hearin for discussions on this work, and Chihway Chang for suggestions on an earlier version of this analysis and draft. We also thank the \textsc{Camels} collaboration for making their data publicly available and easily accessible to the community. DA is supported by NSF grant No. 2108168.

All analysis in this work was enabled greatly by the following software: \textsc{Pandas} \citep{Mckinney2011pandas}, \textsc{NumPy} \citep{vanderWalt2011Numpy}, \textsc{SciPy} \citep{Virtanen2020Scipy}, and \textsc{Matplotlib} \citep{Hunter2007Matplotlib}. We have also used
the Astrophysics Data Service (\href{https://ui.adsabs.harvard.edu/}{ADS}) and \href{https://arxiv.org/}{\texttt{arXiv}} preprint repository extensively during this project and the writing of the paper.

\section*{Data Availability}

The tabulated \textsc{Kllr} parameters and the interpolator framework used in this work are made available at \url{https://github.com/mufanshao/CAMELS_ScalingRelations}. The public \textsc{Camels} data, including the \textsc{Rockstar} catalogs we use in this work, can be found at \url{https://camels.readthedocs.io}.

\bibliographystyle{mnras}
\bibliography{References}

\label{lastpage}
\end{document}